\documentclass[fleqn,usenatbib]{mnras}
\usepackage[dvipsnames]{xcolor}
\usepackage{newtxtext,newtxmath}
\usepackage[normalem]{ulem}

\usepackage[T1]{fontenc}
\usepackage{ae,aecompl}
\usepackage{times}
\usepackage{graphicx}
\graphicspath{ {./images/} }
\usepackage{amsmath}	
\usepackage{verbatim} 

\def\R500c{R_{\rm 500c}}
\def\R200m{R_{\rm 200m}}
\def\M500c{M_{\rm 500c}}
\def\Msun{M_{\odot}}

\newcommand{\cvir}{c_{\text{vir}}}
\newcommand{\Xoff}{X_{\text{off}}}
\newcommand{\Sgas}{S_{\text{gas}}}
\newcommand{\Sdm}{S_{\text{dm}}}
\newcommand{\Qgas}{Q_{\text{gas}}}
\newcommand{\Qdm}{Q_{\text{dm}}}

\newcommand{\pkg}[1]{\textsc{#1}}

\title[Gas Shapes in Dark Matter Haloes with Interpretable Machine Learning]{SHAPing the Gas: Understanding Gas Shapes in Dark Matter Haloes with Interpretable Machine Learning}

\author[Machado \& Avestruz et al.]{Luis Fernando Machado Poletti Valle,$^1$\thanks{E-mail: luisfernando.machado@aya.yale.edu} 
Camille Avestruz,$^{2,3}$\thanks{E-mail: cavestru@umich.edu} 
David J.\ Barnes,$^{4}$ 
Arya Farahi,$^{5,6}$
\newauthor
Erwin T.\ Lau$^{7}$, 
Daisuke Nagai$^{1}$
\\
$^1${Department of Physics, Yale University, New Haven, CT 06520, U.S.A.}\\
$^2${Department of Physics, University of Michigan, Ann Arbor, MI, 48109, U.S.A.}\\
$^3${Leinweber Center for Theoretical Physics, University of Michigan, Ann Arbor, MI, 48109, U.S.A.}\\
$^4${Department of Physics, Kavli Institute for Astrophysics and Space Research, Massachusetts Institute of Technology, Cambridge, MA 02139, U.S.A.}\\
$^5${The Michigan Institute for Data Science, University of Michigan, Ann Arbor, MI 48109, U.S.A.}\\
$^6${Department of Statistics and
Data Science, The University of Texas at Austin, TX 78712, U.S.A.}\\
$^7${Department of Physics, University of Miami, Coral Gables, FL 33124, U.S.A.}\\
}

\pubyear{2020}

\begin{document}
\label{firstpage}
\pagerange{\pageref{firstpage}--\pageref{lastpage}}
\maketitle

\begin{abstract}
The non-spherical shapes of dark matter and gas distributions introduce systematic uncertainties that affect observable-mass relations and selection functions of galaxy groups and clusters. However, the triaxial gas distributions depend on the non-linear physical processes of halo formation histories and baryonic physics, which are challenging to model accurately.
In this study we explore a machine learning approach for modelling the dependence of gas shapes on dark matter and baryonic properties.
With data from the IllustrisTNG hydrodynamical cosmological simulations, we develop a machine learning pipeline that applies \pkg{XGBoost}, an implementation of gradient boosted decision trees, to predict radial profiles of gas shapes from halo properties. 
We show that \pkg{XGBoost} models can accurately predict gas shape profiles in dark matter haloes. We also explore model interpretability with \pkg{SHAP}, a method that identifies the most predictive properties at different halo radii. We find that baryonic properties best predict gas shapes in halo cores, whereas dark matter shapes are the main predictors in the halo outskirts. This work demonstrates the power of interpretable machine learning in modelling observable properties of dark matter haloes in the era of multi-wavelength cosmological surveys.
\end{abstract}


\section{Introduction}
Upcoming large-scale surveys of galaxy clusters and groups in X-ray, microwave, optical wavelengths (such as eROSITA, CMB-S4, and the Rubin Observatory) will cover large cosmological volumes and provide accurate maps of the gas and dark matter distributions in galaxy clusters and groups, which in turn enable us to improve cosmological and astrophysical constraints \citep[see][for reviews]{allen_etal11, kravtsov_borgani12,pratt_etal19}. 
Baryonic physics dominate systematics in several branches of cluster-based cosmological constraints, such as using the Sunyaev-Zeldovich (SZ) effect measurements of pressure and density profiles \citep[e.g.,][]{amodeo_etal20} and cross-correlations of the thermal SZ and gravitational lensing signals \citep[e.g.,][]{hill_spergel14, van_waerbeke_etal14, hojjati_etal17,osato_etal18,osato_etal20}. 

To maximize the scientific returns of upcoming large-scale surveys, we must understand and model the non-linear physics of baryons in dark matter haloes of groups and clusters \citep[see][for review]{wechsler_tinker18}. 
The relationship between baryons and their host dark matter halo enables us to create mock survey observations, which are required for testing models and validating survey science pipelines. Baryonic physics operates in wide ranges of physical scales, and thus is difficult to self-consistently model in cosmological simulations. This self-consistent modelling is particularly difficult in simulations containing enough cluster-size, high resolution haloes to generate complementary mock data for the next generation of surveys. 

Semi-analytic models have been particularly useful for modeling galaxies in large-scale surveys \citep[see][for review]{somerville_dave15}. By painting stellar properties onto dark matter haloes in relatively cheap dark matter-only cosmological simulations, semi-analytic models offer more computationally feasible way to explore the parameter space of galaxy formation physics. Existing semi-analytic models of the halo-galaxy connection, however, do not model halo gas, which are important for cluster and group-size haloes for upcoming microwave and X-ray surveys.

One of the common simplifying assumptions in modelling baryons when dark matter haloes is the spherical symmetry in the halo gas distribution. If not properly modeled, this assumption leads to systematic uncertainties that significantly affect observables and halo selection functions of upcoming surveys. Specifically, the spherical assumption does not properly capture the scatter in observable-mass relations of galaxy clusters \citep{buote_humphrey_2012,chen_etal19,ansarifard20}. This scatter can introduce systematic uncertainties in model predictions such as those of multiwavelength cross-correlation of clusters and groups. In addition, observables are sensitive to viewing angle with respect to the semi-major axes of the triaxial shapes of dark matter and gas haloes with respect to the line-of-sight \citep[e.g.,][]{hamana_etal12, dietrich_etal14, osato_etal17}.  Therefore, the shapes of gas in haloes are an important systematic that needs to be incorporated in mock observations for the next generation of multiwavelength surveys. 

Cosmological simulations have shown that dark matter haloes can be described as being triaxial \citep[e.g.,][]{Jing_2002, Allgood_2006}, with the triaxial shapes depending on halo formation history \citep[e.g.,][]{despali_etal14,lau2020correlations}.  Given the non-linear evolution of the shapes of dark matter haloes, and the unconstrained baryonic effects, analytic modeling of dark matter and gaseous halo shapes is inherently challenging. Using the Zeldovich approximation to evolve from the triaxial peaks of the primordial gas density field, \citet{lee_etal05} derived an ansatz of triaxial shapes for dark matter haloes, which however did not match the mass and redshift dependence of triaxial shapes in cosmological simulations, nor did it capture non-linear halo growth (e.g., merger effects). Baryonic effects, which also consist of non-linear effects, affect the shapes of dark matter haloes in a similar manner. Specifically, radiative cooling in the centres of haloes can lead to more spherical halo shapes, due to enhanced condensation of baryons (cold gas and stars in the core regions of the halo), hence deepening the potential well and effectively making the halo more concentrated and spherical \citep{kazantzidis_etal04, debattisa_etal08}.  On the other hand, feedback from the supermassive blackholes in the halo centres can offset the effects of baryon condensation on halo shape by preventing the gas from cooling and forming stars in the core, either by thermal heating or by ejecting the gas from the inner regions \citep{bryan_etal13, suto_etal17,chua_etal19}. Additionally, the hot gas in massive dark matter haloes is expected deviate from hydrostatic equilibrium, since the gas distributions follow the gravitational potential of the underlying dark matter \citep{buote_canizares_1994}, but are modulated by baryonic effects \citep[e.g.,][]{lau_etal11, biffi_etal13}. 
The complex interplay of halo growth, non-linear effects, and baryonic physics makes it challenging to accurately model the underlying gas and dark matter distributions in cosmological simulations using methods such as empirical modelling or multivariate regression. 

To expand on the capabilities of cosmological simulations, without requiring full simulation suites, we intend to develop an {\em effective} model of gas shapes that allow us to forward-model baryonic properties onto large-scale dark matter-only simulations. This approach can be more efficient than developing entire suites of cosmological simulations, without requiring full analytic models. To obtain efficient and accurate models of gas distributions within galaxy clusters, the machine learning approach offers a promising avenue, since it can learn non-linear relationships between features in the data that may be difficult to parametrize. Recent papers have used such machine learning techniques to model the mapping between galaxy and dark matter halo properties \citep{lovell_etal21,elliott_etal21}.

In this paper we use the \pkg{XGBoost} model \citep{chenandguestrin16} to predict gas shapes based on different halo properties. We also adopt the SHAP (SHapley Additive exPlanations) method \citep{Shapley195317AV, NEURIPS2017_7062} in $\S$\ref{sec:results} to improve model interpretability, allowing us to develop insights into the multiple physical processes that influence halo shapes.

This paper is structured as follows. $\S$\ref{sec:simulation} describes the IllustrisTNG cosmological hydrodynamical simulations used in this study.  $\S$\ref{sec:xgboost} motivates and describes our choice in machine learning approaches with the \pkg{XGBoost} model and the \pkg{SHAP} method.  $\S$\ref{sec:results} presents \pkg{XGBoost} model predictions of the radial profiles of gas shape in dark matter haloes, the interpretation of our trained models with \pkg{SHAP} that identifies the most predictive halo properties, and comparison of model performance with Conditional Abundance Matching (henceforth CAM), which is an empirical non-machine learning based approach.  Finally, $\S$\ref{sec:conclusions} summarizes our main findings.

\section{Simulation}\label{sec:simulation}

\subsection{Hydrodynamical Simulations}

The goal of this paper is to determine how well different halo properties can predict the gas shape profile of galaxy groups and clusters. To make predictions, we train a machine learning algorithm on simulated data (see $\S$\ref{sec:xgboost} for details on the algorithm). Most of our analysis and discussions use data from the TNG300 simulation data from the IllustrisTNG project \citep{naiman18,pillepich18a,nelson18a,marinacci18,springel18}. To compare with models trained on a population of lower mass haloes, we also use data from the TNG100 simulation. We illustrate the respective mass distributions in Figure~\ref{fig:m500_distribution}, where we overlay TNG100 and TNG300 halo populations used in this work. 

TNG100 and TNG300 are cosmological simulations created with the \textsc{Arepo} code \citep{springel10} in a periodic box with side lengths 75 Mpc (TNG100) and 205 Mpc (TNG300), and target masses $m_{\text{baryon}} = 1.4 \times 10^6 M_{\odot}$ (TNG100) and $m_{\text{baryon}} = 1.1 \times 10^7 M_{\odot}$ (TNG300). TNG100 and TNG300 were run with a flat cosmology consistent with \citet{planck16XIII}, with $h = 0.6774$, $\Omega_{\Lambda}=0.6911$, $\Omega_{m}=0.3089$, $\Omega_{b}=0.0486$, $\sigma_{8}=0.8159$, and $n_{s}=0.9667$. IllustrisTNG has been shown to yield realistic clusters \citep[e.g.,][]{Barnes2018, vogelsberger18}. For more details regarding the IllustrisTNG simulations, we refer the reader to \citet{pillepich18a}.

\subsection{Halo Catalog Contents}\label{sec:halocat}

For the model training, we use the halo catalog information provided in the publicly available TNG100 and TNG300 datasets. 
For each snapshot, the haloes are identified using the  Friends-of-Friends (FoF) algorithm, wherein all particles (dark matter, stars, blackholes and gas) in each halo are linked together using a standard linking length of $b=0.2$ times the average particle separation, centered on the nearest most-bound DM particle. For each FoF halo, the Spherical Overdensity (SO) halo masses $M_{500c}$, $M_{200c}$, $M_{\text{vir}}$ are computed by summing the mass of all particles and cells enclosed within $R_{500c}$, $R_{200c}$, and $R_{\text{vir}}$, respectively\footnote[1]{$R_{200c}$ (resp. $R_{500c}$) corresponds to the radius of a sphere with density equal to 200 (resp. 500) times the critical density of the Universe at the halo's redshift. $R_{\text{vir}}$ refers to the definition from \cite{Bryan_1998}.}. We only consider haloes that are not subhaloes, i.e., haloes that are not residing within $R_{200c}$ of another halo. 

\begin{figure*}
    \centering
    \includegraphics[width=0.47\textwidth]{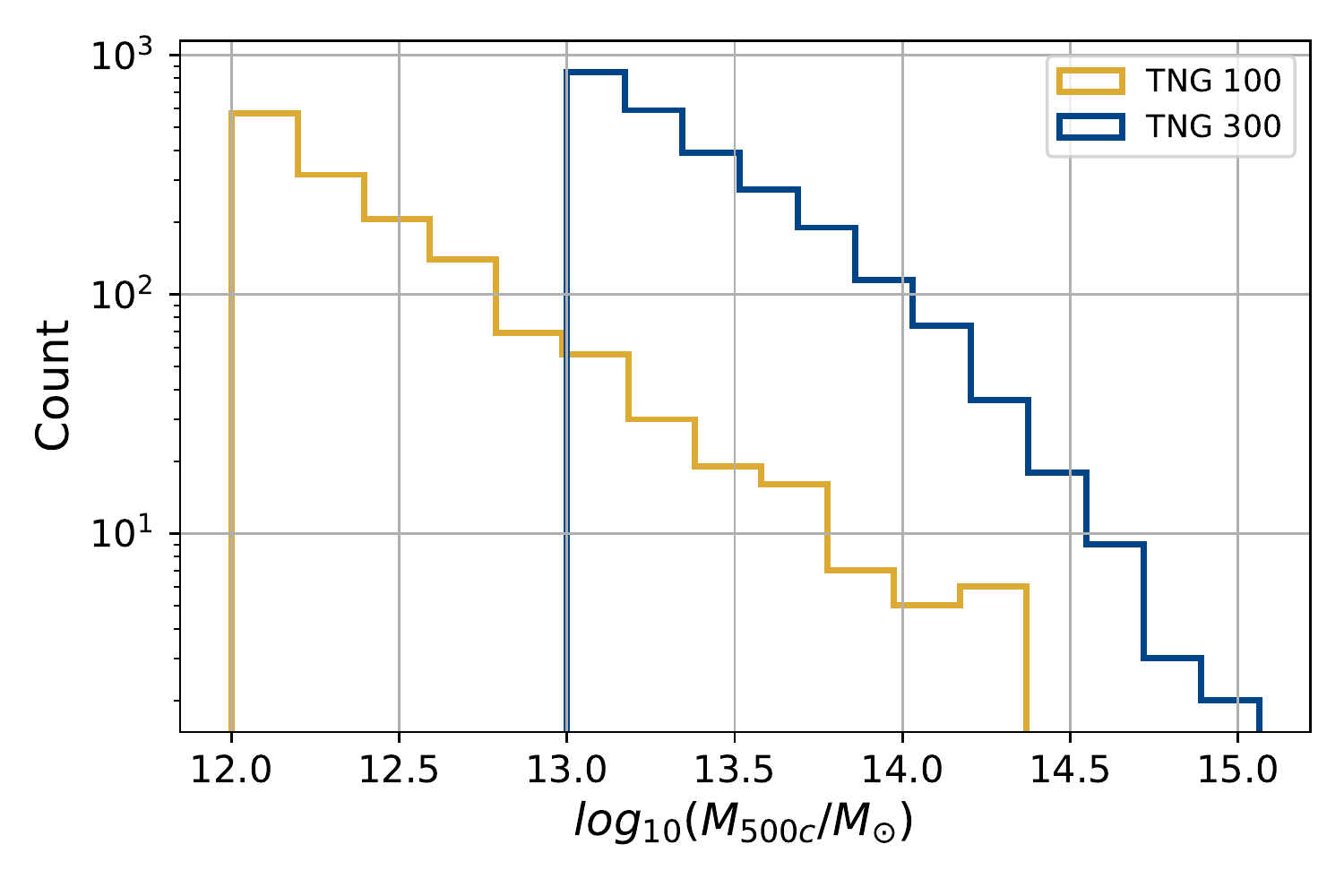}
    \includegraphics[width=0.47\textwidth]{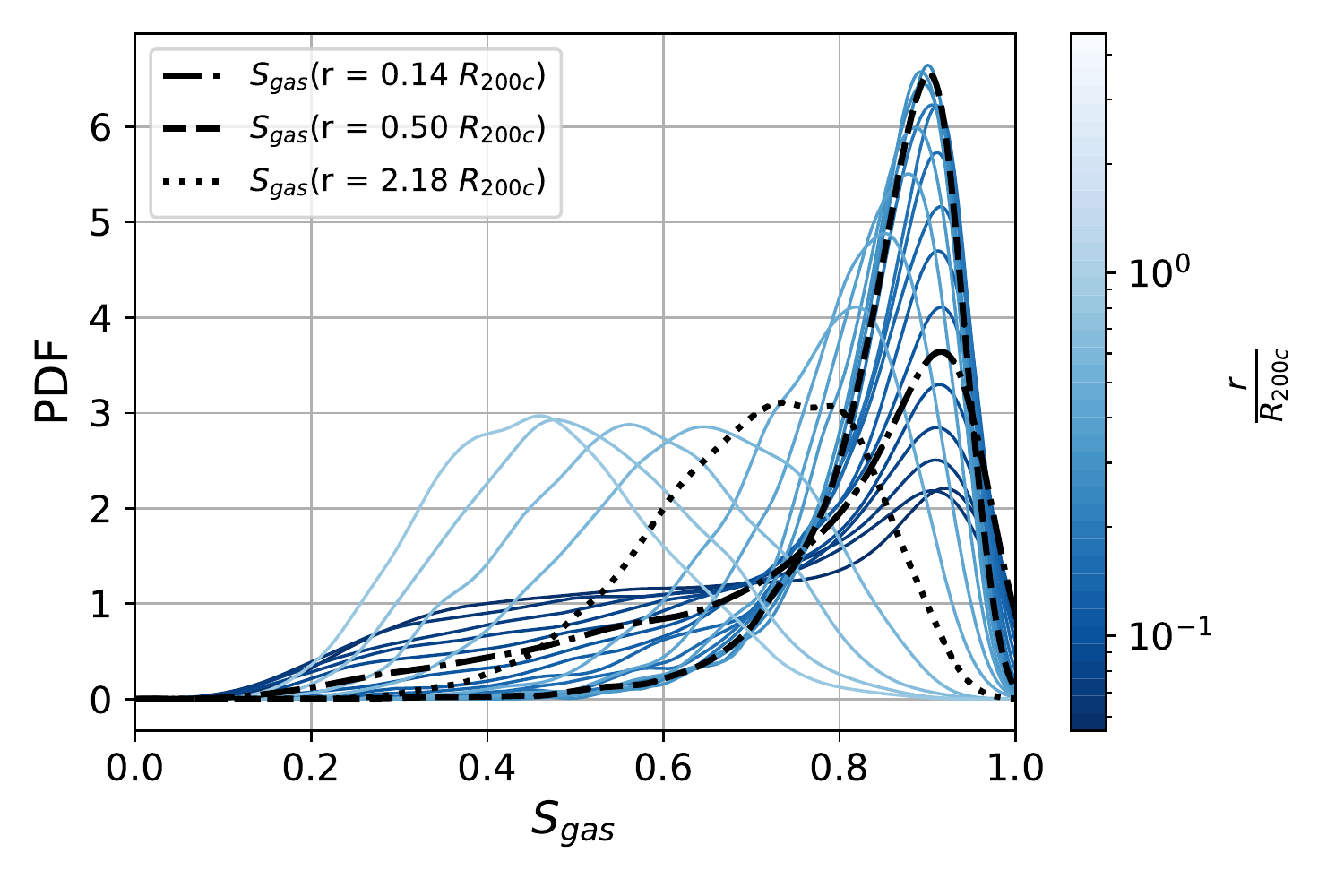} 
    \caption{Datasets included in this study. \emph{Left panel: } Distributions of $M_{500c}$ for the halo populations used in this study from both TNG100 (blue solid histogram) and TNG300 (yellow shaded histogram). For TNG100, we use haloes with $M_{500c} \geq 10^{12} \Msun$, while for TNG300, we use haloes with $M_{500c} \geq 10^{13} \Msun$. The methodology in this paper can also be used to analyze wide halo mass ranges, but we perform separate analyses on each data set to assess the impact of halo masses on learned feature importance in $\S$\ref{sec:feature_ranking}. \emph{Right panel:} Distributions of $\Sgas$ for the TNG300 haloes, color coded by the radius at which we measure gas shape. The distributions shift from mostly spherical (i.e., $\Sgas \sim 1$) at the core to less spherical in the outskirts. We indicate the inner, intermediate, and outer radial bins 
    using dotted, dashed, and dot dashed,  respectively (see $\S$ \ref{sec:results_distributions} for discussions on the radial binning).}
    \label{fig:m500_distribution}
\end{figure*}

Figure \ref{fig:m500_distribution} shows the range of halo masses covered in the population, which includes a total of 2548 haloes. To train the machine learning model, we include the density profiles for gas ($\rho_{\text{gas}}(r)$), stars ($\rho_{\text{stars}}(r)$), and dark matter ($\rho_{\text{dm}}(r)$) within each halo. In addition to mass properties, we compute the localized triaxial shape parameters at different radial annuli, corresponding to the minor-to-major ($S$) and middle-to-major ($Q$) axis ratios, as described in $\S$\ref{sec:shape_parameters}. Furthermore, we include halo formation proxies such as the halo concentration $\cvir$, the peak-centroid offset $X_{\text{off}}$, and the dark matter surface pressure $P_{\text{dm}}$ (measured within the spherical shell between $[0.8 ,1.0]\times R_{\text{vir}}$), which are described in $\S$\ref{sec:formation_parameters}. Finally, we include information about the halo accretion history via the mass accretion rate parameter $\Gamma$, as described in $\S$\ref{sec:accretion_parameters}.

The halo data points used in our analysis are summarized in Table \ref{table:halo_catalog}. Note that we use all information from the snapshot at $z = 0$, except for the mass accretion history, which naturally requires information from earlier snapshots.

\begin{table*}
\centering
\begin{tabular}{c c} 
 \hline\hline
 Group & List of Parameters \\ [0.5ex] 
 \hline
 Halo Properties & $M_{200c}$; $M_{500c}$; $R_{200c}$; $R_{500c}$; $\cvir$; $\Xoff$; $P_{\text{dm}}$; $\rho_{\text{dm}}(r)$ \\
 Dark Matter Shape Profiles & $S_{\text{dm}}(r)$; $Q_{\text{dm}}(r)$ \\
 Accretion History Parameters & $a\left(\frac{M}{M(z=0)}\right)$, $\frac{M}{M(z=0)} \in \{0.5, 0.7\}$; $\Gamma_{200c}(a)$, $a \in \{0.5, 0.6, 0.7, 0.8, 0.9\}$ \\
 Baryonic Density Profiles & $\rho_{\text{gas}}(r)$; $\rho_{\text{stars}}(r)$ \\
 \hline
\end{tabular}
\caption{Halo information used in the machine learning model. Shape profiles are obtained through the iterative inertia tensor procedure described in \S\ref{sec:shape_parameters}. Formation history proxies are described in $\S$\ref{sec:formation_parameters}. Mass accretion rate parameters are determined from tracing data, according to \S\ref{sec:accretion_parameters}. Note that these profiles are interpolated into 25 radial bins, to maintain significant granularity while not overwhelming the learning process.
}
\label{table:halo_catalog}
\end{table*}

\subsubsection{Shape Profiles} \label{sec:shape_parameters}

To compute the shape profile, we adopt a radial binning scheme with 25 bins, spaced log-uniformly between the cluster core and outskirt, spanning the radial range of $0.05 \leq r/R_{200c} \leq 4.5$. We choose 25 bins as a large enough quantity to capture the radial dependence of the shape parameters, while also low enough as to not overwhelm the machine learning process, which becomes increasingly computationally expensive as we increase the number of bins. To illustrate how different physical processes impact gas shapes at different radii, we focus on the shape measurements at three radial bins: $r/R_{200c} \in [0.13,0.15]$, $[0.45,0.55]$, and $[2.0 , 2.4]$, which correspond respectively to the core, intermediate regions, and outskirts of the analyzed clusters. We also compute the global gas shape parameters as measured within $R_{500c}$, which we select since $R_{500c}$ is accessible to most current X-ray observations of the ICM. 

We parametrize gas and dark matter distributions as triaxial ellipsoids with axis lengths $a \geq b \geq c$, and express the shape of such distributions in terms of the axis ratios $S \equiv c/a$ (minor-to-major) and $Q \equiv b/a$ (middle-to-major). Note that, by definition, we always have $S \leq Q$, and that larger values of $S$ and $Q$ imply more spherical distributions. Additionally, note that we can determine shape parameters for different particle types, which in particular gives rise to gas ($\Sgas$, $\Qgas$) and dark matter ($\Sdm$, $\Qdm$) shape profiles.

We follow the iterative method described in \citet{lau_etal11} when computing the axis ratio profiles. The axis ratios are derived from the mass tensor:
\begin{equation}\label{eq:mass_tensor}
\mathcal{M}_{ij} \equiv \frac{1}{\sum^N_p m_p}\sum^N_p m_p x_i x_j,
\end{equation}
which is computed by starting with particles with mass $m_p$ within a given radial shell $[r,r+dr]$, where $x_i$ corresponds to the Cartesian coordinate ($x_i \in x, y, z$) of the particle relative to the halo centre. 
The major, intermediate, and minor axes $(a,b,c)$ of the triaxial ellipsoid are computed as the square roots of the sorted eigenvalues of the resulting mass tensor. If the ellipsoidal (or elliptical) radius of the particle, $r_{\rm ep} = a\sqrt{(x'/a)^2 + (y'/b)^2+(z'/c)^2}$ (with $x',y',z'$ being the coordinates of the particle in the frame of eigenvectors of the mass tensor) lies within the radial shell $[r,r+dr]$, the particle is then included in the next computation iteration of the mass tensor. 
The iteration continues until the axis ratios $Q$ and $S$ converge to some predetermined accuracy (in this work, $<5\%$). To ensure that the shape measurements are not affected by subhaloes, we remove particles in subhaloes when computing the mass tensor.

In $\S$\ref{sec:results_cam_comparisons}, we compute the global axis ratios $Q$ and $S$ for all particles within $R_{500c}$. We select $R_{500c}$ since it is the outermost radius where we can reliably measure ICM properties with current X-ray instruments 
\citep[see section 2 of][for motivation]{Vikhlinin_2009}. In that case, we follow the same procedures described above, but instead of requiring the particle's ellipsoidal radius to lie within the radial shell $[r, r+dr]$ and the ellipsoidal radius to lie inside $R_{500c}$ at each iteration step.

We note that there is no unique way of estimating halo shapes. There are other methods in the literature for computing the mass tensor. Some uses all particles interior to the radius instead of within a shell \citep[e.g.,][]{bryan_etal13}. Some methods use a $r_{\rm ep}^{-2}$ weighting scheme \citep[e.g.,][]{dubinski_carlberg91, Allgood_2006, schneider_etal12} to reduce effects of substructures; others use the spherical radius $r_{\rm sph} = a\sqrt{x'^2 + y'^2+z'^2}$ instead when selecting the particles, often without iterations \citep[e.g.,][]{cole_lacey96, bailin_steinmetz05}. Other methods calculate the shape of density contours by selecting particles based on their local density values \citep[e.g.,][]{Jing_2002}. 
Compared to these other shape measurement methods, our method has been shown to produce accurate halo shape as function of radius \citep{zemp_etal11}.

\subsubsection{Halo Formation Proxies}\label{sec:formation_parameters}

The halo formation history significantly impacts the gas contents within haloes, and can also alter the gas distribution shapes encoded by $S_{\text{gas}}$. As a result, we include commonly used halo formation proxies as inputs in the machine learning model, to assess whether the model can find correlation between halo formation history and the final observed gas shapes in haloes. We include the following halo formation proxies and morphological metrics:

\begin{itemize}
    \item $X_{\text{off}}$ - Distance offset between the peak total mass density and the centre of mass for the total mass distribution. Large values of $X_{\text{off}}$ indicate high merger activities and a less relaxed halo environment.
    \item $\cvir$ - Defined by $\cvir \equiv R_{\text{vir}}/R_s$, where $R_s$ is the scale radius found by fitting the DM density profile to the Navarro, Frenk and White (NFW) profile \citep{nfw96}. High halo concentration correlates with earlier formation epochs, which in turn indicate that a halo has had more time to relax.
    \item $P_{\text{dm}}$ - Dark matter surface pressure, computed by including all particles within the spherical shell between $[0.8,1.0] R_{\text{vir}}$, as defined as $P_{\text{dm}} \equiv 1/(3V) \sum_i m_i v_i^2$ for DM particle $i$ within the shell and $V$ is the shell volume \citep[see Equation 4 in][]{Shaw_2006}.
\end{itemize}

\subsubsection{Mass Accretion Rate} \label{sec:accretion_parameters}

To compute the mass accretion rate, we use the merger trees computed with \pkg{Sublink} \citep{Rodriguez-Gomez15}. The \pkg{Sublink} algorithm identifies the progenitor sequences across snapshots by identifying haloes that share the most bound particles between consecutive snapshots. This results in mass accretion histories for every halo in the population. From the mass accretion history $M(z)$, we measure the mass accretion rate of each galaxy cluster at a given time as:
\begin{equation}
\Gamma(t) = \frac{\log_{10}(M(t)/(M(t - t_{\text{dyn}}))}{\log_{10}(a(t))/(a(t-t_{\text{dyn}}))},
\end{equation}\label{eq:gamma}
modified from the original definition of $\Gamma$ in \citet{diemerkravtsov15}.  Here, we use the mass definition of $M = M_{200c}$ from the halo catalog described in \ref{sec:halocat}. The dynamical time, $t_{\text{dyn}}$, is given by:
\begin{equation}
t_{\text{dyn}}(z) = 2^{3/2} \cdot t_{\text{H}} \cdot \left(\frac{\rho_{\Delta}}{\rho_{\rm crit}}\right)^{-1/2} = 2^{3/2} \cdot t_{\text{H}} \cdot \Delta^{-1/2},
\end{equation}\label{eq:tdyn}
where $M_{\Delta}$ is defined as the mass enclosed in a cluster-centric radius at which the average density, $\rho_{\Delta}$, is $\Delta$ times the critical density of the universe at that time, $\rho_{\rm crit}(z)$.  We express $t_{\text{dyn}}(z)$ in terms of the Hubble time, $t_{\text{H}} \equiv 1/H(z)$. For $\Delta = 200c$, this gives $t_{\text{dyn}} \approx t_{\text{H}}/5$. Note, we modify the dynamical timescale in the denominator of the original use of $\Gamma$ in \citet{diemerkravtsov15} to probe timescales more closely connected to large-scale changes in the matter distribution. For instance, see example in \citet{chen_etal19} illustrating how different timescales for accretion rate measurements change the correlation strength between the accretion rate and cluster gas shapes.

We note that this analysis could be performed to obtain values for $\Gamma$ at any selected radius. We choose to focus on $\Gamma_{200c}$ since $R_{200c}$ contains the outer regions of haloes, and therefore variations in $M_{200c}$ reflect more accurately the broader merger activity of the halo. We also include mass accretion history parametrised by $a\left({M}/{M(z=0)}\right)$, the expansion factor when the halo mass was some fraction of its present-day value, $M/M(z=0)$.  For each halo, we calculate the expansion factor when the halo is $0.5$ and $0.7$ of its present-day value, respectively $a\left(0.5\right)$ and $a\left(0.7\right)$.  We use these as additional features in the ``Accretion History Parameter'' group of features to train the models. Note that both $\Gamma_{200c}$ and $a\left({M}/{M(z=0)}\right)$ encode instantaneous rates of mass accretion, but cover different timescales and provide a more complete picture of the mass evolution in the sampled haloes.

\section{Interpretable Machine Learning Model}\label{sec:xgboost} 

Through the use of a machine learning model, we aim to both make predictions and interpret those predictions through a physical lens. We employ \pkg{XGBoost} to predict gas shape based on different halo properties. We then determine how the model made its predictions, by employing the \pkg{SHAP} method to quantify the relative importance of each feature in the final model prediction. We describe the motivation and pipeline for using \pkg{XGBoost} and \pkg{SHAP} in the following subsections.

\subsection{Regression with Machine Learning}

In this work, the primary goal is to develop a method to predict a continuous value (gas shape) based on the value of one or several predictor values (halo properties), which is a straightforward {\it regression analysis}. Among many machine learning models to perform such regression analyses, we chose \pkg{XGBoost}, a scalable model that has been vetted in multiple applications within the machine learning community.

\pkg{XGBoost} is an implementation of gradient boosted decision trees, a set of machine learning techniques used for regression and classification \citep{chenandguestrin16}. We note that \pkg{XGBoost} performs significantly faster and more accurately than similar gradient-boosted tree solutions. We refer the reader to \citet{gradientboostedguide} for the theoretically-backed design choices in its underlying optimization algorithms.

The crux of this paper is to accurately predict the gas shape profile for a galaxy group and cluster based on a given set of halo parameters. We train the \pkg{XGBoost} regressor model using properties from a sample population of galaxy groups and clusters.  This population is divided into two random samples, training and validation, to allow for a proper evaluation of the resulting trained models.  We also select different groups of halo properties to serve as input training features.  The output target values are comprised of the gas shape (e.g., the ratio of the semi-major axis to the semi-minor axis, $S_{\text{gas}}$) measured at a given halo-centric radius for each galaxy group or cluster in the training set. Note, we train a separate \pkg{XGBoost} model for each radius at which we want a shape prediction. We also note that the target data consists of values ranging between 0 and 1, and thus we must be careful in verifying that final model predictions satisfy this constraint. The trained models enable both a prediction and a ranking of relative feature importance.  To assess the model performance, we evaluate each model on another subset of our galaxy groups and clusters (the {\it validation set}), predicting the gas shape from properties of galaxy groups or clusters that were not included in the training process.  By applying \pkg{SHAP}, we can also determine which features have the most impact in the final shape prediction at any given radius.

\subsection{Interpretability via SHAP Method}\label{sec:interpretability}

While machine learning models are powerful in making accurate prediction, explaining how the model reached a specific prediction is a difficult task. Our goal is not just to make the most accurate prediction, but also to understand the patterns picked up by our machine learning model in order to inform our physical intuition of the dependencies between gas shapes and halo properties.

We employ the \pkg{SHAP} (SHapley Additive exPlanations) method in order to explain the trained models. \pkg{SHAP} follows a game-theoretic approach to explain the model predictions by computing the relative contributions of each feature to the final predictions \citep{Shapley195317AV}. Specifically, \pkg{SHAP} values represent the changes in the expected model prediction when conditioning on that feature \citep{NEURIPS2017_7062}. \pkg{SHAP}'s strong theoretical guarantees regarding the \emph{accuracy} and \emph{consistency} of the estimated feature contribution distinguishes this tool from other interpretability methods. We employ \pkg{SHAP} method in conjunction with the trained \pkg{XGBoost} model to interpret the trained models.

\subsection{Data Pre-Processing and Rescaling}

The study of galaxy cluster properties naturally brings about a challenge with the scales involved in the different observable data fields. In fact, the set of feature inputs presented in Table \ref{table:halo_catalog} contains values ranging from unit scales (e.g., $\Sdm$) to several orders of magnitude larger (e.g., $M_{200c}$). Machine learning models, in particular \pkg{XGBoost}, are highly sensitive to features at different scales, and tend to erroneously bias towards using the larger feature values as decision points, even if the underlying data does not correlate as strongly with the target field. To prevent this issue, and obtain meaningful impact from every feature in the dataset, we first perform a common pre-processing of the input data. This involves applying a logarithm to all fields in the input dataset, and then shifting each feature's distribution to one ranging from 0 to 1. As a result, every data field maintains their original sorting properties, while also allowing \pkg{XGBoost} to focus on the meaningful correlation between features.

Machine learning models also suffer from the risk of \emph{overfitting}, whereby a model learns highly non-linear trends in the input parameter space, and fits its underlying model to predict the training target data very accurately, at the expense of accuracy when presented with new data. A common approach to prevent overfitting is the use of a training-test data split, in which a subsample from the entire available dataset is used for the training process, and the remaining data is saved for a validation step once the training is complete. This process allows us to verify that the model performs well with both the training and the test datasets, which indicates the model has not suffered from overfitting. In this study, we apply the commonly used 80-20 split, with 80\% of the halo population is used for training, and the remaining 20\% is used in the validation step.

\subsection{Hyperparameter optimization}\label{sec:hyperparameters}

Machine learning models often have parameters that control the learning process, and such \emph{hyperparameters} are not related to the actual feature data used during model training. These hyperpameters are fundamental in creating generalizable and accurate models, while preventing overfitting of the training data. As a first step in training such models, it is essential to perform hyperparameter optimization, the process through which we select optimal learning parameters before proceeding with the actual model training. Each hyperparameter controls a separate configuration in the learning process, with the most commonly noted as the main drivers of accuracy improvements being:

\begin{itemize}
    \item Learning Rate: Controls the rate at which new trees are added to the model, in order to account for new information. Lower values lead to a slower addition of new trees, which slows down the learning process and prevents model overfitting.
    \item Number of Estimators: Number of boosted trees fitted via the training process. Larger values lead to more accurate models, at the risk of overfitting to the training data.
\end{itemize}

We perform a traditional grid search, via which every combination of the considered hyperparameters is used to train a different model on the same sample feature set. The models are compared via a mean squared error loss function (RMSE), and we determine the optimal set of hyperparameters as the set that minimizes the RMSE on the sample training set. Consider a machine learning model trained on $N$ haloes at different radial bins. The $i$-th halo at the $j$-th radial bin contains target data $S_{{\rm gas},i,j}$, and predicted value (from the ML model) $\hat{S}_{{\rm gas},i,j}$. Then, the RMSE value for this model is computed as:
\begin{equation}
    {\rm RMSE}(r_j) = \frac{1}{N} \sum\limits_{i=1}^N \left(S_{{\rm gas},i}(r_j) - \hat{S}_{{\rm gas},i}(r_j) \right)^2,
\end{equation}
and similarly for $\Qgas$. 

We also determine the relative improvement on the final RMSE values gained from each hyperparameter. We performed this grid search at 3 different radii, to compare optimal hyperparameters in the inner, intermediate, and outer regions of the haloes. We note that the optimal hyperparameter values have a weak dependence on the radial bin selected, with the final improvements in RMSE varying less than 5\% across all sampled radial bins. Therefore, we select the most common set as the optimal set of hyperparameters throughout our analysis.

The set of considered hyperparameters, ranked by largest relative improvement in the final RMSE, is described in Table \ref{table:hyperparameter_space}. We note that the average RMSE obtained with the optimal hyperparameters showed slight improvements (by $\sim 20\%$) over similar models trained with randomly selected hyperparameters.

\begin{table*}
\centering
\begin{tabular}{c c c} 
 \hline\hline
\pkg{XGBoost} Hyperparameter & Explored Values & Optimal Value \\ [0.5ex] 
 \hline
 Learning Rate & $10^{-5}, 10^{-4}, 10^{-3}, 10^{-2}, 10^{-1}, 1, 10$ & $10^{-1}$ \\
 Number of Estimators & $10, 50, 100, 300, 500, 700, 1000$ & $300$ \\
 Maximum Depth & $3, 4, 5, 6, 7$ & $5$ \\
 Minimum Child Weight & $3, 4, 5, 6$ & $4$ \\
 Gamma & $0, 0.1, 0.2, 0.3$ & $0$ \\
 Subsample & $0.5, 0.6, 0.7, 0.8, 0.9$ & $0.8$ \\
 Column Sample Rate By Tree & $0.4, 0.5, 0.6, 0.7, 0.8$ & $0.6$ \\
 Alpha & $10^{-9}, 10^{-7}, 10^{-5}, 10^{-3}, 10^{-2}, 10^{-1}, 1, 100$ & $10^{-2}$ \\
 \hline
\end{tabular}
\caption{List of hyperparameters searched when preparing the model, ranked by overall impact in the final model accuracy. The individual hyperparameters are described in $\S$\ref{sec:hyperparameters}. We note that \emph{learning rate} and \emph{number of estimators} provide over 90\% of the combined improvement in model accuracy. Furthermore, the set of optimal hyperparameters only varied minimally across radial bins, and as a result we apply the same optimal set across all models, to maintain the consistency and limit the impact to the different feature sets under consideration.}
\label{table:hyperparameter_space}
\end{table*}

\subsection{Interpretability via Feature Ranking}\label{sec:feature_ranking}

We train \pkg{XGBoost} models using different subsets of features as the training data sets. The different subsets combine different entries from Table \ref{table:halo_catalog}, and all models include minimally the \emph{Halo Properties} subset. We compare the model accuracy for each feature set as a function of radius, to determine which set of features contribute the most in determining the gas shape distributions in our halo population. In addition to the comparison across feature sets, we explore the relative importance of different features within a given feature set in order to determine which features contribute the most to the final models. We intend to quantify the predictive power of different halo properties on baryonic shape parameters. We then employ \pkg{SHAP} method to rank features based on their overall contribution to the predictions. Specifically, we employ \texttt{TreeSHAP} algorithm \citep{lundberg2020local2global}, which has been developed for estimating \pkg{SHAP} values using tree-based machine learning models such as \pkg{XGBoost} model. Features with large absolute average SHAP values contribute more to the prediction and hence more important. 

The computed \pkg{SHAP} values for every trained model allow us to determine the most important features for every iteration of the \pkg{XGBoost} models, which in turn inform our physical intuition of the dependencies between gas shapes and several halo properties at varying radial ranges, from the core to the outer regions of groups and clusters.

We compared the relative feature importance determined from the \pkg{XGBoost} built-in feature ranking methods. We then followed previous comparisons between different implementations of \texttt{TreeSHAP} algorithm and converged on applying the ``interventional'' method of \texttt{TreeSHAP}, ensuring that it yields consistent results when compared to the best feature importance ranking methods from \pkg{XGBoost} \citep{chen2020true}.

\section{Results}\label{sec:results}
\subsection{Relationship between Shape Distribution and Halo Properties}\label{sec:results_distributions}

\begin{figure*}
    \centering
    \includegraphics[width=\textwidth]{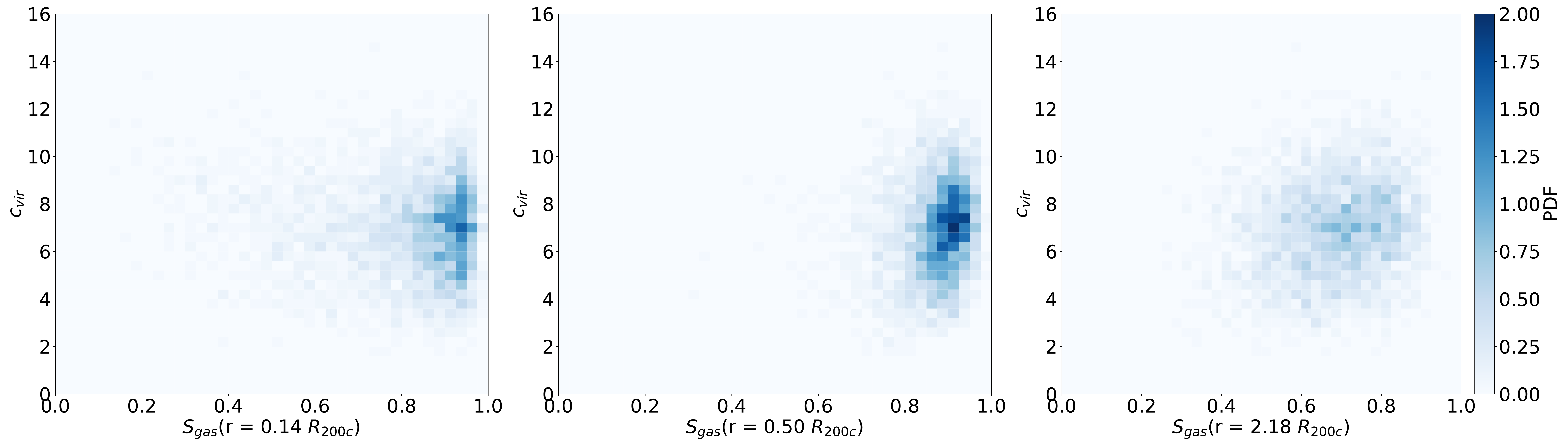}
    \includegraphics[width=\textwidth]{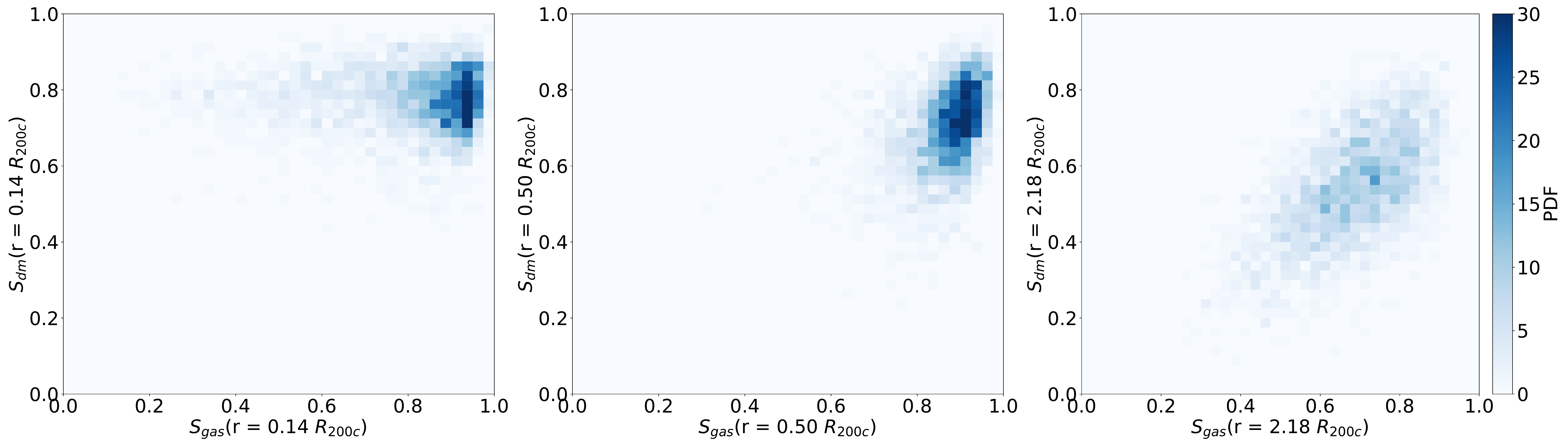}
    \caption{
    Example 2-dimensional distributions of haloes in different parameter spaces. We consider these distributions at 3 different radial bins: inner ($0.13  < r/R_{200c} < 0.15$, \emph{left}), intermediate ($0.45  < r/R_{200c} < 0.55$, \emph{middle}), and outer ($2.0 < r/R_{200c}  < 2.4$, \emph{right}). The {\em top row} shows distribution in $\cvir$ vs. gas shape space, while the {\em bottom row} shows the distribution in dark matter shape vs. gas shape space, with dark matter shapes measured at the same radial bin as the gas shape. The halo distributions change as a function of radial bin, which demonstrates that the high dimensional space of halo properties, as well as their correlations, is a complex space to capture with traditional statistical methods.
    }
    \label{fig:sgas_binned}
\end{figure*}

Before applying \pkg{XGBoost} models to study the sampled haloes, we first illustrate how the distribution of gas shapes measured at different radii varies with different halo properties.  As an example, the panels of Figure \ref{fig:sgas_binned} show the 2-dimensional histograms of the gas shape axis ratio, $S \equiv c/a$, measured at three different radii, namely: inner ($r/R_{200c} \in [0.13,0.15]$, \emph{left}), intermediate ($r/R_{200c} \in [0.45,0.55]$, \emph{middle}), and outer ($r/R_{200c} \in [2.0 , 2.4]$, \emph{right}). We describe the radial binning choices in $\S$\ref{sec:shape_parameters}.  We select these sample radial bins based on the premise that different physical processes may impact the gas at different radii. 
We expect that baryonic effects such as cooling and energy feedback impact the gas in the core regions, whereas mass accretion processes affect gas distributions in the outskirts.  

In Figure \ref{fig:sgas_binned}, the top panel shows gas shape axis ratio distributions of objects in the space of halo concentration, $\cvir$, and gas shape measurements at example inner (left), intermediate (middle), and outer (right) radial bins.  The bottom panel shows the distribution in the space of dark matter shape axis ratio, $\Sdm$, at that same radial bin. 
Comparing the three columns, we first note that the distribution of $\Sgas$ connects with rounder gas shapes at inner and intermediate radii, with most haloes between $0.8 < \Sgas < 1.0$.  On the other hand, in the outer radial bin, there are more haloes with lower values of $\Sgas$. This result generally agrees with previous studies \citep{Samsing_2012, Lau_2012}, and indicates that gas distributions are less spherical, on average, in the outskirts than they are in the inner regions.  

These example distributions illustrate that different properties correlate differently with gas shapes measured at different radii.  Concentration, often used as a secondary halo parameter to predict baryonic physical properties, does not trace gas shapes for massive objects from the TNG300 simulation.  However, in $\S$\ref{sec:results_shap_mass_dependence}, we show how concentration is in fact more important for lower mass objects from TNG100.  Dark matter shapes correlate most strongly with gas shapes at the halo outskirts, but present weaker correlations towards the halo cores.  In this work, we aim to create a predictive machine learning model that captures the combined information from an arbitrary number of properties in order to predict gas shapes at different radii. It is natural to imagine that a non-linear combination of properties might best predict gas shape, with such combination parameters depending on the particular radial bins under analysis.  We use \pkg{XGBoost}, a machine learning algorithm, to perform this task. The next sections present the predictive performance of \pkg{XGBoost} models (see $\S$\ref{sec:featuresubsets}) and how we can leverage interpretable machine learning models to determine the most predictive halo properties at every radial bin (see $\S$\ref{sec:model_accuracy} and $\S$\ref{sec:results_most_important_features}).

\subsection{Impact of feature subsets on model performance}\label{sec:featuresubsets}

In this subsection, we describe the performance of the \pkg{XGBoost} models trained with different subsets of halo properties. We selected 7 different sets of halo properties (\emph{feature sets}), which are different combinations of the parameters described in Table \ref{table:halo_catalog}. Specifically, we trained a \pkg{XGBoost} regressor to predict gas shapes at each of the 25 radial bins available using each of 7 feature subsets.  The targets for model predictions are $S_{\text{gas}}$, the localized gas short-to-major axis ratio at the different radial bins.  We discuss summarized results for a total of $7 \times 25 = 175$ regressor models.  

For each trained model, we compute the RMSE between the predicted values and the actual (target) values. We perform this computation for the reserved test sets, in order to prevent overfitting and to measure the final accuracy of our model for different radial bins and different feature set inputs.

\begin{figure*}
    \centering
    \includegraphics[width=0.99\textwidth]{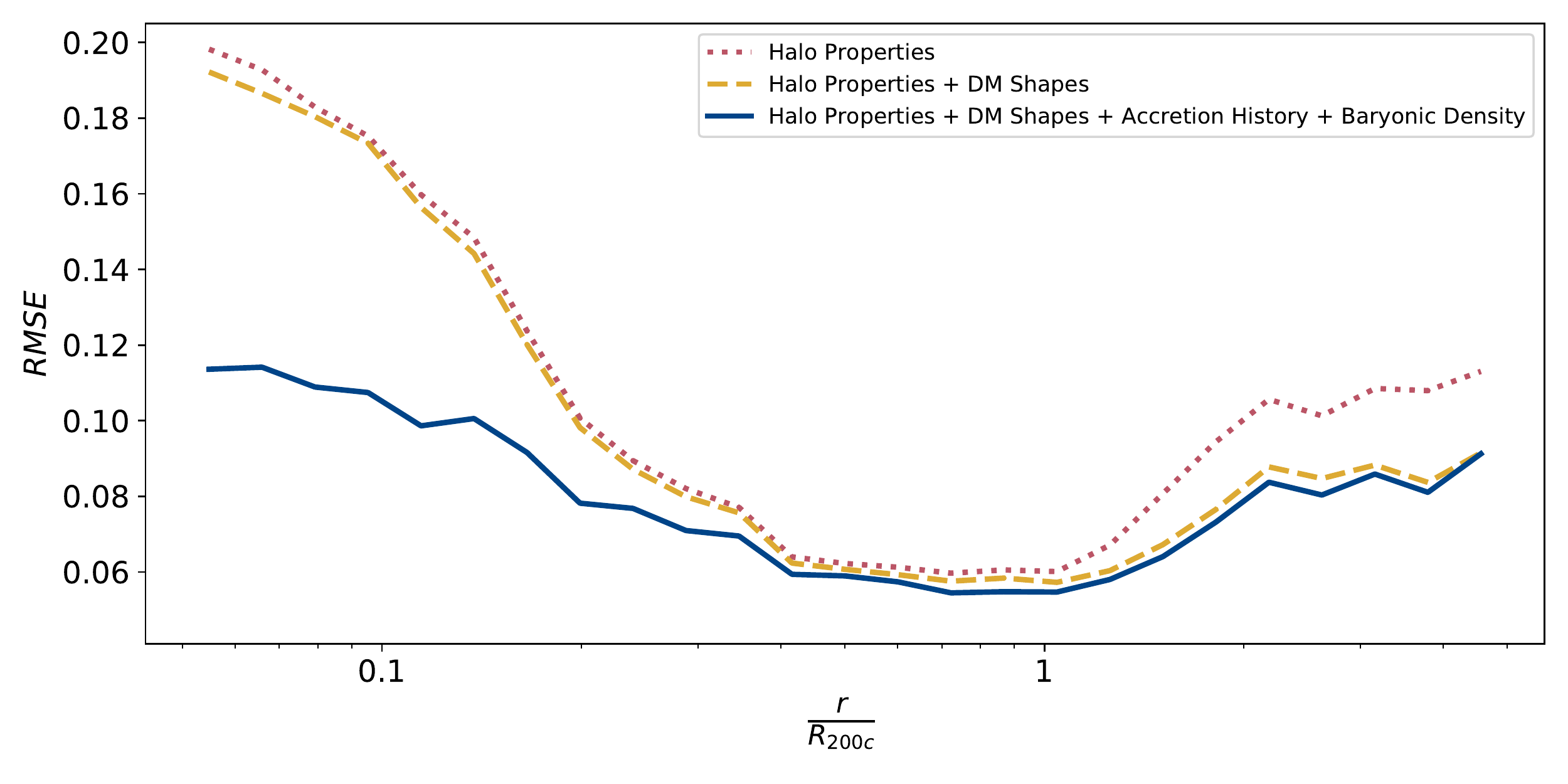}
    \caption{Accuracy of models quantified by the root mean squared error, RMSE, as a function of the cluster centric radius of the gas shape prediction normalized by $R_{200c}$.  Each line corresponds to a different feature set used in model training.  Lower values of the RMSE indicate more accurate models. Models trained with more information systematically perform better at all radii, but the addition of baryonic information improves model performance the most for gas shape predictions in the innermost radii (dark blue solid and cyan dashed lines).  The models trained to predict outskirt gas shapes with the feature set of only dark matter information with dark matter shape have comparable accuracy to the models trained with baryonic information.}
    \label{fig:rmses_feature_sets_ratios_test}
\end{figure*}

Figure \ref{fig:rmses_feature_sets_ratios_test} illustrates the RMSE of models we trained, as a function of the radius at which the model predicts gas shape.  Lower values of the RMSE indicate higher model accuracy.  Each line corresponds to models trained with a given feature set labeled in the legend.  The feature sets correspond to subsets of properties listed in Table~\ref{table:halo_catalog} and described in $\S$\ref{sec:feature_ranking}. All models have an intersection of features that consist of parameters listed in the first row corresponding to the group ``Halo properties''.  The order of feature subsets listed in the legend of Figure~\ref{fig:rmses_feature_sets_ratios_test} in the order of increasing information content of the feature sets; the dotted red line shows model accuracy for the baseline feature set with only {\it Halo Properties}, the dashed yellow line for models with dark matter shapes included, and the dark solid blue line illustrated model performance with the maximal feature set that includes information of accretion history and baryonic density information.  Note, we include an illustration of relative model accuracies of all feature sets we test in the Appendix Figure \ref{fig:appendix_rmses}, but limit our discussion of model accuracy to these three feature sets, which had the most significant variations in behavior.  The inclusion of more features in model training generally improves model accuracy.  In particular, models trained with all available features including baryonic information systematically best-perform at all radii.  However, model predictions outside the cluster cores, $r/R_{200c}\gtrsim 0.15$, have less significant improvement with additional features.  

For example, our predictions in the innermost radial bin have an improvement in RMSE of almost a factor of 2 when we include baryon density profile information. This is to be expected, since gas and stellar distributions at the inner regions are largely affected by baryonic effects such as feedback and cannot be accurately determined solely from information about the dark matter distribution.  We further discuss this in $\S\ref{sec:results_most_important_features}$. The model performance of models trained with {\it Halo Properties + DM Shapes} and models trained with baryon density information becomes comparable when predicting gas shapes at larger radii, with minimal difference in RMSEs for models that predict gas shapes in the range $0.4\lesssim r/R_{200c}\lesssim1$.  In fact, at $r=0.5 R_{200c}$ the relative RMSEs of the dark matter-only models are only $10\%$ away from the RMSE obtained with all of the available input data. As we move to even larger radii, $r/R_{200c}$, the better performing models are those that were trained with features that include dark matter shape information, and baryonic information does not significantly further improve model accuracy predictions of gas shape outside of $r/R_{200c} \geq 0.4 $.  However, models without dark matter shapes (e.g., \emph{Halo Properties} shown in the dotted red line) perform relatively worse in these regions.  Combined, these results indicate that dark matter shapes are a powerful and necessary predictor of gas shapes beyond $0.5R_{200c}$. Physically, this observation is consistent with outer regions of haloes being largely determined by their dark matter distributions, and less affected by baryonic effects \citep{Nagai_2007, chan_etal15, Schaller_2015}.

The relative model performance comparison indicates that baryonic density profiles are the main predictors of gas shape in the core, while dark matter shape is an important predictor of gas shape in the outskirts.  We also note that accretion history information added relatively little improvement to model accuracy in the outskirts; this was also the case in absence of DM Shapes (see Appendix \ref{sec:appendix_rmse}), counterintuitive to what one might expect.  Results for $\Qgas$, the semi-minor axis ratio gas shapes, are similar. We do not include results from the analysis here, but more \pkg{SHAP} figures can be found in Appendix \ref{sec:appendix_100_300_comparison}.

Finally, we add the caveat that more information does not necessarily lead to better models, with the effects dependent on the machine learning algorithm.  For example, redundant features can lead to reduced model accuracy if the machine learning method relies on small neighborhoods of feature space to make predictions.  Also, irrelevant features can reduce model accuracy for algorithms such as decision tree based methods.  Here, decisions deeper in the tree may depend on randomness introduced by irrelevant features leading to an overfitting of training data \citep[for more details on the impact of additional data on machine learning models, see][]{nakkiran2019data, advani2017highdimensional}.


\begin{figure*}
    \centering
    \includegraphics[width=\textwidth]{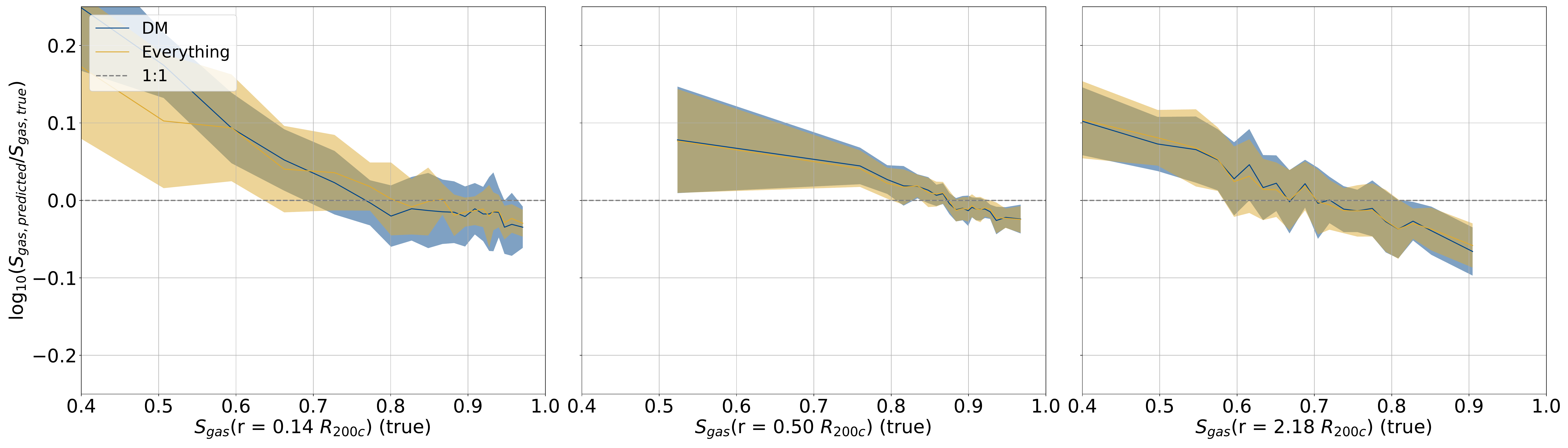}
    \includegraphics[width=\textwidth]{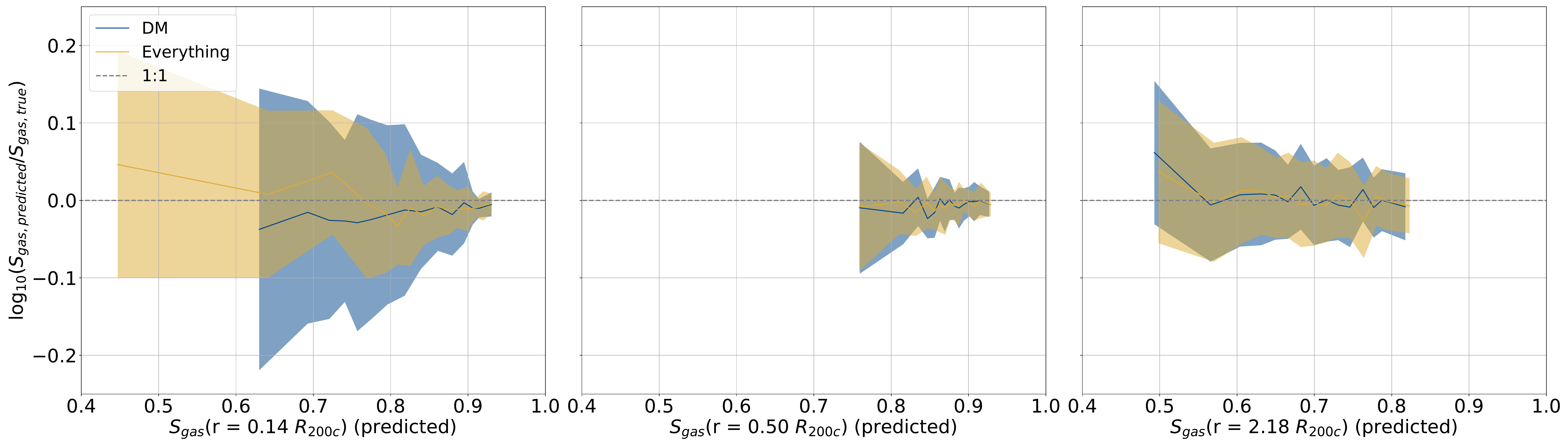}
    \caption{Predictive performance of the \pkg{XGBoost} models at characteristic inner, intermediate, and outermost radial bins, with performance quantified by the $log_{10}$ of the ratio between the predicted shapes to true gas shapes.  Top row shows the performance as a function of the {\it true} gas shapes at that radius, and the bottom shows the performance as a function of the {\it predicted} gas shapes at that radius.  Panels correspond to radial bins: $r/R_{200c} \in [0.13,0.15]$ (\emph{left}), $r/R_{200c}  \in [0.45,0.55]$ (\emph{middle}), $r/R_{200c} \in [2.0 , 2.4]$ (\emph{right}). The solid line indicates the median value of the predictive performance for the \emph{DM} model (blue) and the \emph{Everything} model (orange) as described in the legend for Figure \ref{fig:rmses_feature_sets_ratios_test}). A perfectly accurate model corresponds to a horizontal line at 0.0. The shaded regions indicate the 1$\sigma$ log-normal scatter around the median line. Both models become more accurate at the larger radii.  Model performance also becomes more similar at larger radii.}
    \label{fig:predicted_vs_true_scatter_plots}
\end{figure*}

\subsection{Comparison of model accuracies at different radii}\label{sec:model_accuracy}

In this subsection, we further examine the performance of \pkg{XGBoost} in different radial regions of our sample haloes. From Fig~\ref{fig:rmses_feature_sets_ratios_test} and as discussed in \S~\ref{sec:featuresubsets}, the predictive power of \pkg{XGBoost} noticeably improves the inner regions ($r/R_{200c}\leq 0.4$) with the inclusion of baryon density profiles.  The RMSEs are almost identical for models that predict gas shapes in the inner regions using information from dark matter only information, even when including {\it DM Shapes}.  Models that predict gas shapes in the core only improve with the addition of baryon density information.  In the radial range of ($0.4 \leq r/R_{200c}\leq 1 $), the performance gaps between different \pkg{XGBoost} models are minimal. The models including only dark matter training data present closer accuracy to the models including baryon density information.  This shift suggests that the baryonic data is less necessary for predictions of gas shapes towards halo outskirts. 

We illustrate the trends of the predictive performance of \pkg{XGBoost} models at different radial bins in Figure \ref{fig:predicted_vs_true_scatter_plots} in terms of the $log_{10}$ of the ratio between the predicted and true gas shapes, quantifying the bias in model predictions.  The left, middle, and right panels show the respective performance models predicting gas shapes in a characteristic inner ($r/R_{200c} \in [0.13,0.15]$), intermediate ($r/R_{200c}  \in [0.45,0.55]$), and outer ($r/R_{200c} \in [2.0 , 2.4]$) radial bins.  We use bins consistent with those used in Figure~\ref{fig:sgas_binned}.  The top panel shows the performance as a function of {\it true} gas shape, and the bottom panel as a function of {\it predicted} gas shape.  The blue solid line shows the median $log_{10}$ ratio for the models trained with the {\it DM} model at a fixed shape bin and the orange the same for the {\it Everything} model, where models differ based on subsets of features used during training as described in $\S$\ref{sec:feature_ranking}. 

The top panel illustrates a clear bias, where predictions are high for low values of true gas shape, and predictions are low for high values of true gas shape. This effect is typical for regression problems, where machine learning algorithms exhibit biased predictions at the extrema of the available dataset used in training (e.g., Figure~3 from \citealt{ntampaka_2019} and Figure~8 from \citealt{ho_2019}). 

From the top panel alone, one might assume that we can apply a correction to the model output that removes such a bias with respect to fixed true gas shape.  However, the bottom panel illustrates that there is no such systematic bias at fixed predicted shape for either model.  In the bottom panel, we see that the bias plotted against the predicted gas shape values satisfies the following.  First, the bias is symmetrically distributed, with median values close to the middle of the y-axis that corresponds to the 1:1 line of $S_\text{gas,predicted} = S_\text{gas,true}$
Second, the spread is clustered to a fairly narrow range, with $|\log_{10}(S_\text{gas,predicted}/S_\text{gas,true})|\lesssim0.2$ in the inner radial bin, $|\log_{10}(S_\text{gas,predicted}/S_\text{gas,true})|\lesssim0.1$ in the intermediate bin, and $|\log_{10}(S_\text{gas,predicted}/S_\text{gas,true})|\lesssim0.15$ in the outermost bin (all within $1\sigma$).
Additionally, similar results appear when performing the analysis in Figure \ref{fig:predicted_vs_true_scatter_plots} for $\Qgas$ (not shown) instead of $\Sgas$. The predictive performance follows a similar trend as the one described above, indicating that the model accuracy in predicting gas shapes does not vary greatly with the specific gas shape parameters being predicted. This is an important advantage of applying such training procedures, as they allow one to predict different shape parameters with comparable accuracy and therefore create a more complete model of the gas distributions in the analyzed haloes. 

One detail to note is that the bottom left panel shows that these models exhibit heteroscedasticity, where the variance in error decreases as the predictions go from low values of $\Sgas$ to high values of $\Sgas$. This trend may indicate that there could be additional room for improvement in our model, or simply that there is larger intrinsic stochasticity for low values of $\Sgas$. In particular, this trend is partially expected due to the smaller number of subhaloes with low values of $\Sgas$ at the core. The \pkg{XGBoost} models improve in accuracy with larger sample sizes, and thus the fewer subhaloes with low $\Sgas$ are expected to result in a higher variance in the final predictive performance of the trained models.

An improved model could result through either a transformation of feature and target variables and/or through incorporating additional features that are not in our dataset.  We have checked the general distribution of variables, and all are roughly described by a symmetrical, bell-shaped curve.  If there is additional room for model improvement, it would most likely be through the inclusion of missing variables that better correlate with the stochasticity of physical processes that impact the gas shape, therefore decreasing the variance in the bias.  One such example to improve shape predictions in the core might be properties of the stellar population that indicate recent baryonic feedback events.  On the other hand, there may simply be a level of intrinsic scatter that cannot be tied to additional properties that can be extracted from the simulation.

We also see differences between the {\it DM} and {\it Everything} model performance, primarily in the inner radial bin.  From the upper left panel, we see that both models systematically overpredict gas shapes for haloes with low values of true gas shape.  The difference between the blue and orange lines in that panel shows that the inclusion of baryonic information helps decrease the bias in model predictions at fixed true gas shape.  At intermediate and outer radii, however, the model performance does not significantly change with additional features. 

Specifically, both models improve in accuracy and become more similar to one another towards larger radii. This agrees with the results from Figure \ref{fig:rmses_feature_sets_ratios_test}, where we show the model accuracy as a function of radius, demonstrating that dark matter information is the most predictive halo data set outside the halo core $r/R_{200c} > 0.5$. Adding baryonic information does not significantly increase the model accuracy outside the core. 
On the other hand, adding baryonic information improves the accuracy of the predicted gas shapes in the core radius, especially for lower values of true gas shape of $\Sgas \sim 0.4$, as the median of the $log_{10}$ ratio between predicted and true $\Sgas$ values decreases from $\sim 0.25$ to $\sim 0.18$. At higher true $\Sgas$ values, however, adding baryonic information does not significantly improve the prediction accuracy.

\subsection{Most Predictive Halo Properties}\label{sec:results_most_important_features}

\begin{table*}
\centering
\begin{tabular}{c c c c} 
 \hline\hline
 Radial Range ($R_{200c}$) & \emph{Everything} model & \emph{DM} model & \emph{DM minus DM Shape} model \\ [0.5ex]
 \hline
 $0.05 - 0.2$ & Halo Mass, $\rho_{\text{gas}}$ & Halo Mass, $\Sdm$ & Halo Mass, $\Gamma$ \\
 $0.2 - 0.8$ & $\rho_{\text{gas}}$, DM Shape & DM Shape, $\rho_{\text{dm}}$ & $\rho_{\text{dm}}$, $\Gamma$ \\
 $0.8 - 2$ & $\Xoff$, DM Shape & $\Xoff$, DM Shape & $\Xoff$, $\Gamma$ \\
 $2 - 3$ & DM Shape, $P_{\text{dm}}$ & DM Shape, $P_{\text{dm}}$ & $P_{\text{dm}}$, $\Gamma$ \\
 \hline
\end{tabular}
\caption{Most important features for the \pkg{XGBoost} models at different radial ranges. The feature importances are measured for two feature set models, \emph{Everything}, \emph{DM} and \emph{DM minus DM Shapes}. The \emph{Everything} model gains most of its predictive power from the gas density profiles, which explains the better accuracy towards inner radii, where astrophysical effects are more pronounced. The \emph{DM} model relies on halo mass information to predict gas shape in the inner radii, but gives more weight to DM shape and DM density for the outer regions. In the intermediate regions, there is also contribution from the mass accretion rate information, though it plays a secondary role in the predictions.}
\label{table:most_important_features}
\end{table*}

\begin{figure*}
\centering
\centering
\includegraphics[width=0.47\textwidth]{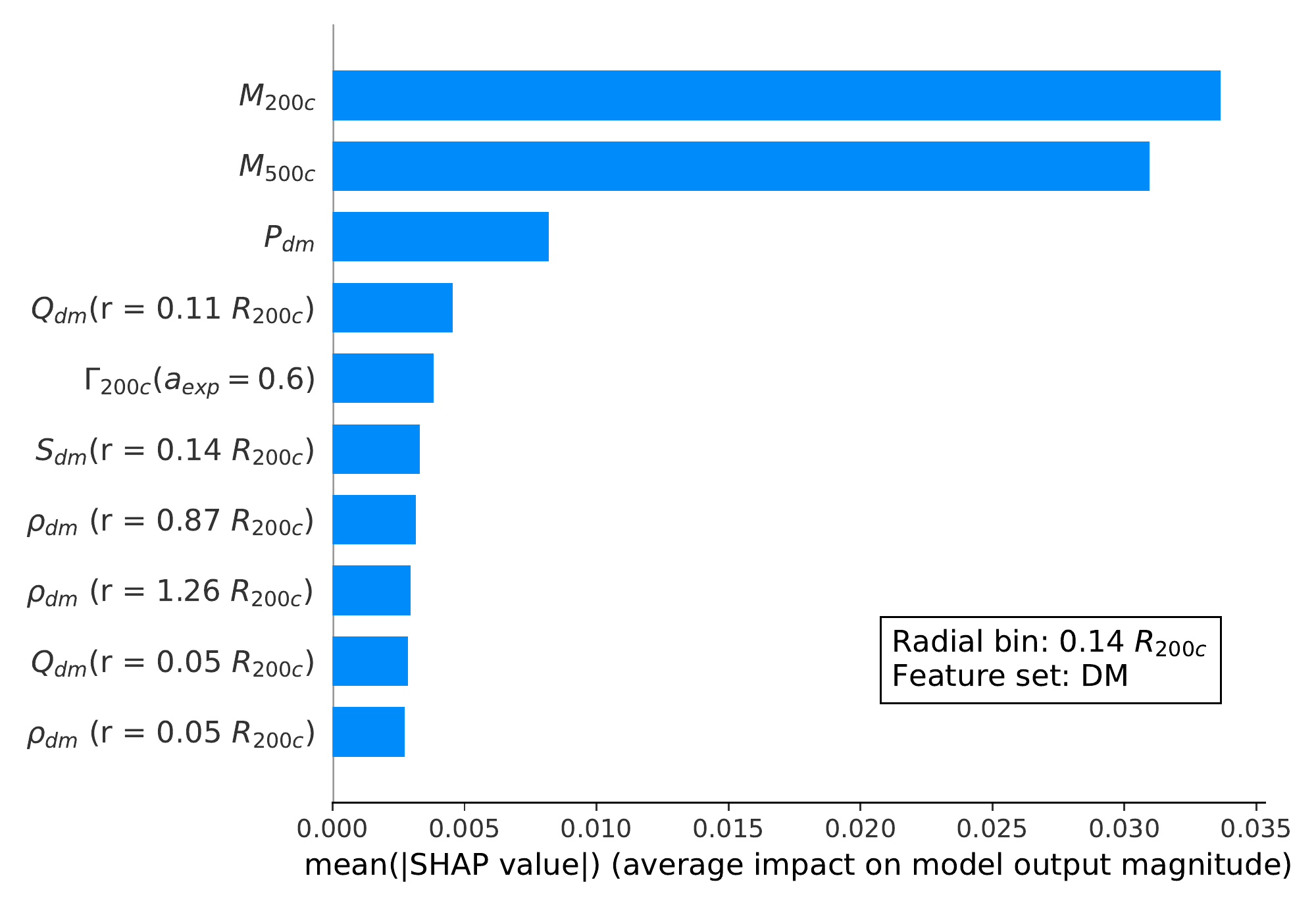}
\includegraphics[width=0.47\textwidth]{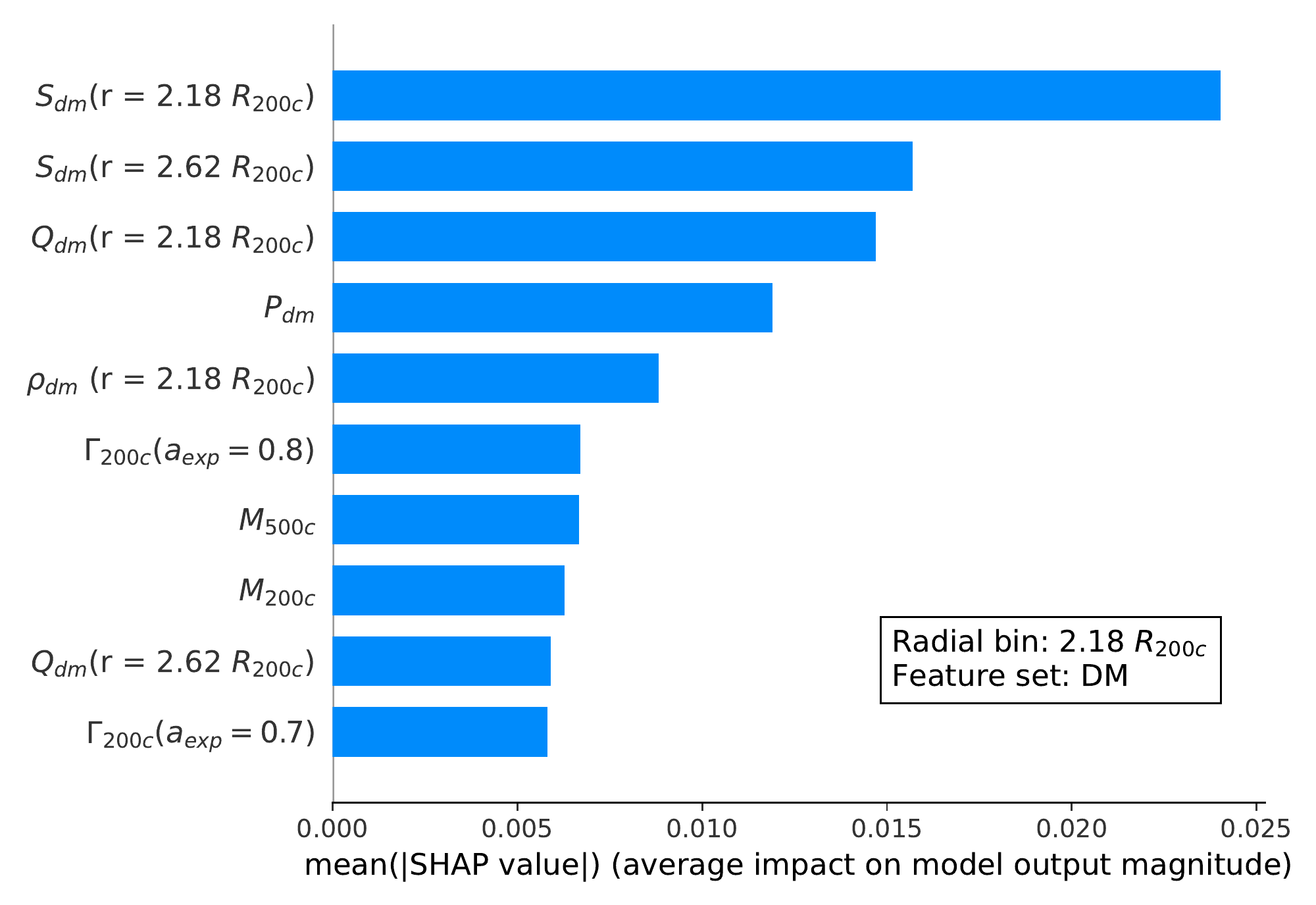}
\caption{Example feature importance rankings determined by \pkg{SHAP} for two radial bins, \emph{left:} inner ($r=0.14 R_{200c}$), and \emph{right:} outer ($r=2.18 R_{200c}$) in TNG300 data. In both models, we use the dark matter (DM) only feature set to train the \pkg{XGBoost} models and predict gas shape $\Sgas$ at the corresponding radial bins. Each figure shows the 10 most predictive features from the selected feature set, ranked by the \pkg{SHAP} algorithm described in $\S$\ref{sec:results_most_important_features}. In particular, global halo masses and localized DM shapes are the strongest DM-only predictor of gas shapes in inner bins, while $\Sdm$ is the strongest predictor of $\Sgas$ in the outer regions of haloes.}
\label{fig:shap_TNG300}
\end{figure*}

We now focus on the relative feature importance rankings obtained from applying \pkg{SHAP} to \pkg{XGBoost} models. Note, we have a separate trained model corresponding to each feature subset predicting each gas shape target at every radial bin.  This means we have 7 feature sets $\times$ 25 radial bins of gas shape, so 175 models.  

For any given \pkg{XGBoost} model, \pkg{SHAP} determines the relative feature importances in predicting the target gas shape values. The feature importance rankings are obtained from the \texttt{TreeSHAP} algorithm described in $\S$\ref{sec:feature_ranking}. Figure \ref{fig:shap_TNG300} shows example \texttt{TreeSHAP} results for the DM feature set at two different radii, the left for gas shape predictions in the core and the right for gas shape predictions in the outermost radial bin. The analysis of all 175 models is summarized in Table \ref{table:most_important_features}.

When trained with on the {\it Everything} feature subset, the \pkg{XGBoost} model training incorporates information from all available properties in its training, including baryonic properties.  With baryonic properties available, we find that $\rho_{\rm gas}$ is one of the most predictive features for gas shape in the radial ranges of $r/R_{200c}\in[0.05,0.2]$.  The other most predictive features these respective radial ranges are the halo mass and underlying dark matter shape.  At larger radii, baryonic properties are not as predictive of the gas shape {\it despite} their inclusion in the model training.  For the radial ranges of $r/R_{200c}\in[0.8,2]$ and $r/R_{200c}\in[2,3]$, the most predictive features are  $X_{\rm off}$, the dark matter shape, and $P_{\rm dm}$, the dark matter surface pressure.

When removing baryonic properties from the features used to train the model, we have the {\it DM model}.  Here, the most important features in the radial ranges of $r/R_{200c}\in[0.8,2]$ and $r/R_{200c}\in[2,3]$ are {\it still}  $X_{\rm off}$, the dark matter shape, and $P_{\rm dm}$.  This result is consistent with what we see in Figure~\ref{fig:rmses_feature_sets_ratios_test}, which illustrates that the addition of baryonic information to model training does not significantly improve the gas shape predictions outside of the cluster core. 

Finally, the right-most column of Table~\ref{table:most_important_features} shows the top two most important features for a models trained with all available dark matter properties except for the underlying dark matter shape profile.  For the {\it DM model} column, the dark matter shape is one of the most powerful predictive features for gas shape for all radii outside of the cluster core, illustrating that the gas distribution largely traces the underlying dark matter distribution at these radii.  This is consistent with the azimuthally averaged analysis from \citet{lau_etal15}.  In absence of dark matter shape information, the model identifies the mass accretion rate, $\Gamma$, as one of the most important features.  However, the mass accretion rate does not provide as much predictive power as we might have expected at large radii.  We checked relative model accuracies (see Appendix \ref{sec:appendix_rmse}; in absence of dark matter shape, models trained with accretion history information show only marginal improvement when compared to models trained with the baseline {\it Halo Properties} feature set.

We can identify physical interpretations that explain some of the feature importance rankings that the \pkg{SHAP} algorithm outputs.  For example, with the {\it Everything model}, localized gas densities $\rho_{\text{gas}}$ are the most predictive feature for localized gas shapes in the radial range of $0.05\leq r/R_{200c}\leq0.8$. The radial gas density information is closely tied to the overall distribution of the gas contents in haloes. From the \pkg{SHAP} results, we can see that the \pkg{XGBoost} algorithm has learned this relationship. 

On the other hand, the dark matter shape is a more powerful predictor of gas shape in radii outside of $r/R_{200c}>0.8$ for both the {\it Everything} and {\it DM} models. The right hand side of Figure~\ref{fig:shap_TNG300} shows the \pkg{SHAP} visualization of relative feature importance in the TNG300 haloes for \pkg{XGBoost} models trained with the dark matter properties, corresponding to the {\it DM model} summary in Table~\ref{table:most_important_features}.  This figure illustrates that the dark matter short-to-major axis ratio, $\Sdm$, is the most powerful predictor of $\Sgas$ in cluster outskirts. Furthermore, the relative importance of the middle-to-major axis ratio, $\Qdm$, is considerably smaller. We note that $\Sgas$ in cluster outskirts is largely driven by filamentary accretion in the outer regions of galaxy clusters, which are mainly defined by one preferential alignment direction \citep[e.g.,][]{gouin21}.  The overall importance of dark matter shape in models that predict gas shape in the outskirts is to be expected, since the gas contents follow the underlying gravitational potential, which in turn is largely determined by the dark matter distribution.

\subsubsection{Predictive properties as a function of halo mass}\label{sec:results_shap_mass_dependence}

In this study we mainly focus on modelling gas shapes in massive haloes from the TNG300 simulations, whose mass range encompasses galaxy groups and clusters (see the orange shaded histogram in Figure~\ref{fig:m500_distribution}). However, the most important features that predict gas shapes in these massive objects may not be the same as those that predict gas shapes in the galaxy scale. In this subsection, we discuss the most predictive properties for models trained in different, narrower mass bins. For this comparison, we sample low mass haloes from TNG100 as well as higher mass haloes from TNG300.

To combine datasets from TNG100 and TNG300, it is essential that (i) the two simulation suites contain similar enough objects, and that (ii) \pkg{XGBoost} and \pkg{SHAP} can determine consistent feature importance rankings across the two simulations. As described in $\S$\ref{sec:simulation}, the IllustrisTNG simulations use similar physical parameters and produce consistent halo populations, which guarantees condition (i). For (ii), we compare results from TNG100 and TNG300 haloes, both sampled from the same mass bin: $\log_{10}(M_{500c}/\Msun) \in [13,14]$, and determine that the \pkg{SHAP} results are consistent across the two simulations. This comparison is described in detail in Appendix~\ref{sec:appendix_100_300_comparison}. The combination of (i) and (ii) allows for the combined use of TNG100 and TNG300 haloes when analyzing the mass dependence of \pkg{SHAP} feature rankings.

To demonstrate how the most predictive features vary as a function of halo mass, we compare \pkg{SHAP} outputs at different mass bins and different radial bins. In particular, we use the dark matter only datasets to predict $\Sgas$ at the \emph{inner} ($r/R_{200c} \in [0.13, 0.15]$) and \emph{outer} ($r/R_{200c} \in [2.0, 2.4]$) radial bins described in $\S$\ref{sec:results_distributions}. We focus on 5 different mass ranges: $\log_{10}(M_{500c}/\Msun) \in [12,12.5]$, and $[12.5,13]$ for TNG100 and $\log_{10}(M_{500c}/\Msun) \in [13,13.5]$, $[13.5,14]$ and $[14,15]$ for TNG300.

For each mass range and radial bin, we train a \pkg{XGBoost} model and apply \pkg{SHAP} to the resulting trained model, similar to the procedure described in $\S$\ref{sec:feature_ranking}. We then compare the resulting \pkg{SHAP} scores for the features that are the most predictive in at least one of the selected mass bins. We note that, due to different effects of baryonic physics at different mass scales, we expect the feature rankings to vary considerably across the selected mass ranges.

\begin{figure*}
    \includegraphics[width=0.48\textwidth]{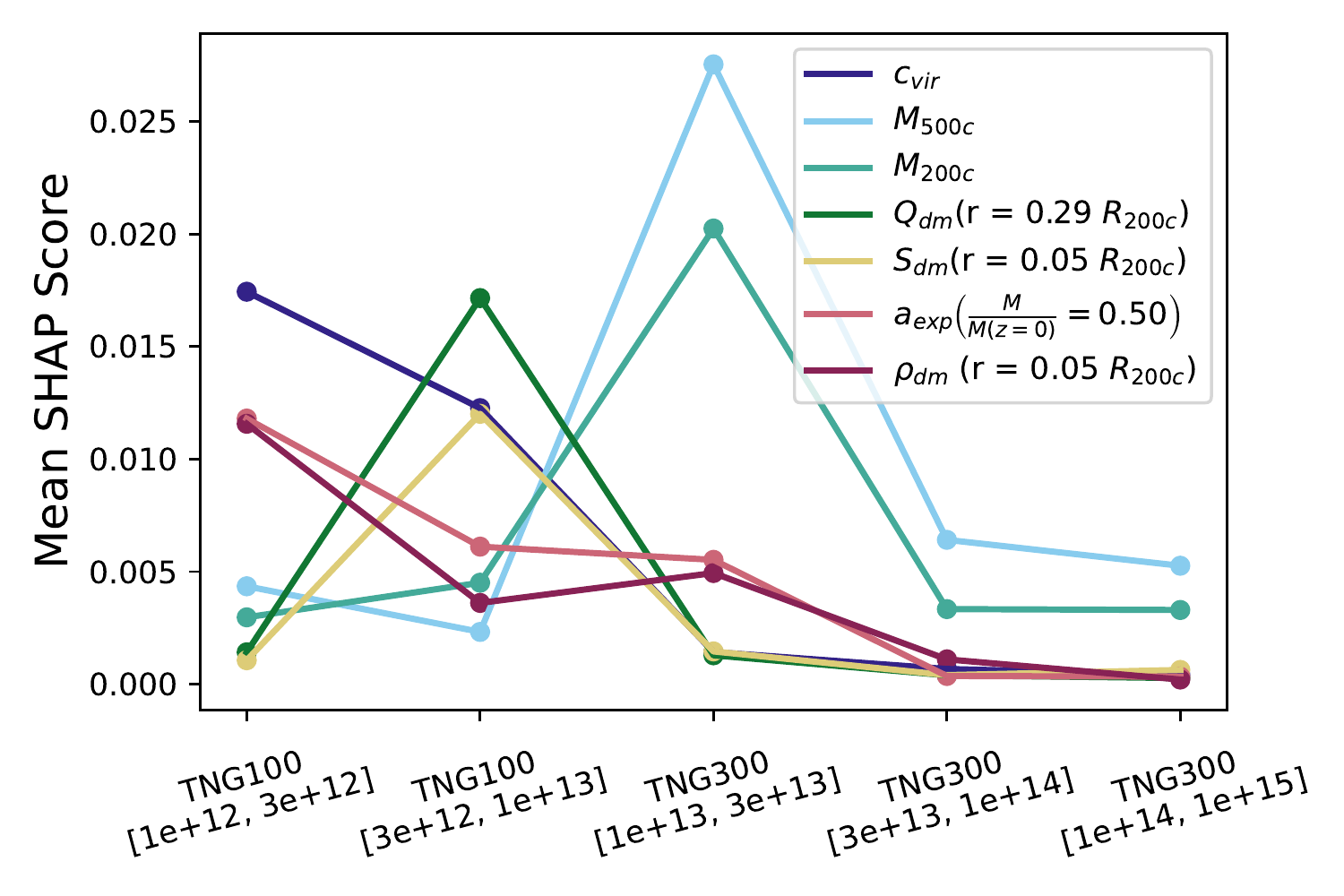}
    \includegraphics[width=0.48\textwidth]{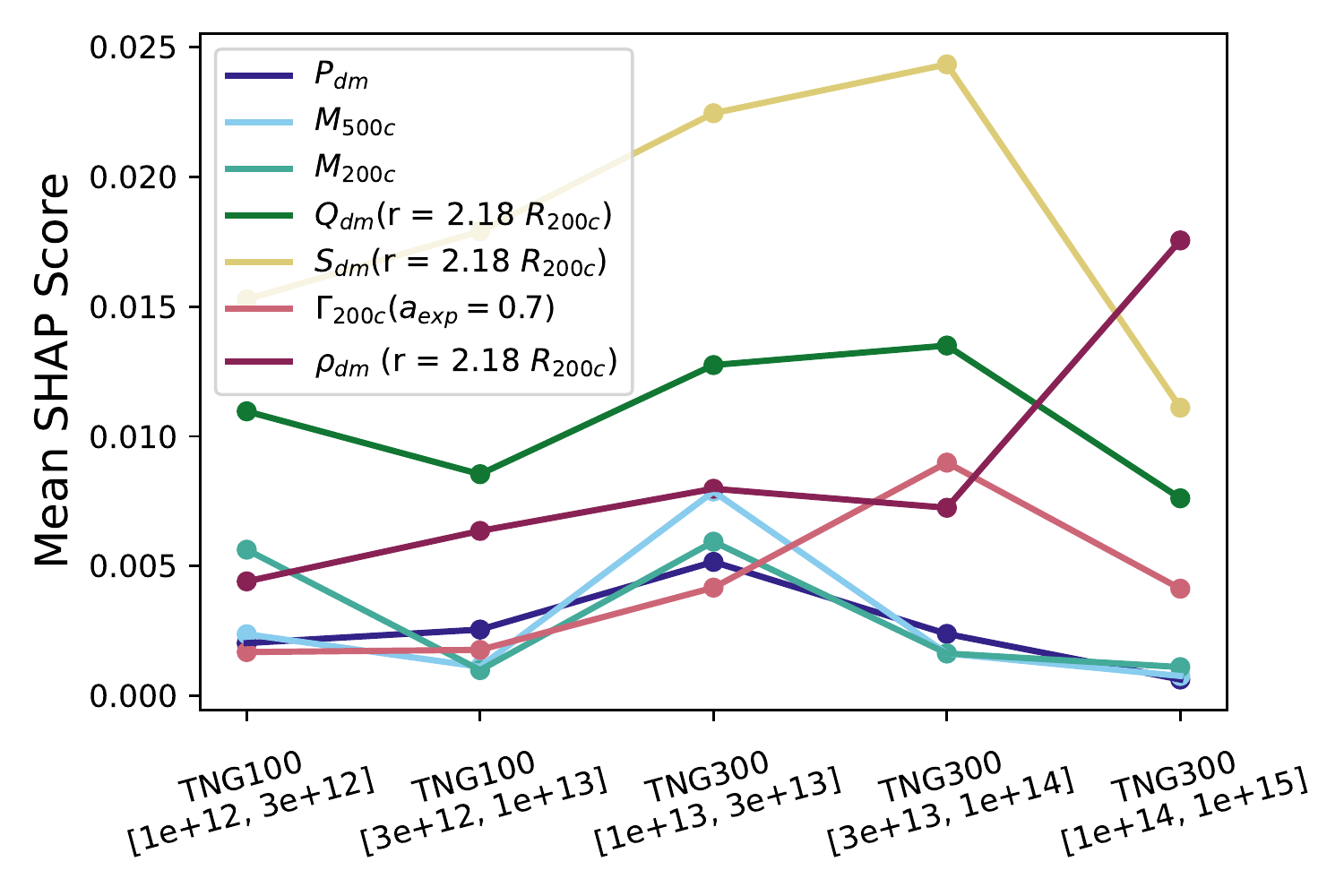}
    \caption{Mass dependence of feature importance scores. We compute \pkg{SHAP} scores for models trained on subsets of the data within 5 different mass bins.  Each mass bin contains either TNG100 or TNG300 haloes. \emph{Left} panel shows the feature importance for models trained to predict $\Sgas$ in the inner radial bin $r/R_{200c} \in [0.13,0.15]$.  \emph{Right} panel shows the same values, for models trained to predict $\Sgas$ in the outer radial bin $r/R_{200c} \in [2.0,2.4]$, as described in $\S$ \ref{sec:results_most_important_features}. The \pkg{SHAP} scores correspond to the same measure as those shown in Figure \ref{fig:shap_TNG300}, with larger values indicating a more predictive feature within any given halo mass range. In halo cores (\emph{left panel}), halo concentration $\cvir$ is the most predictive halo property in lower mass haloes, suggesting different impacts of baryonic physics in the evolution of core gas shapes in the lower mass haloes. In the halo outskirts (\emph{right panel}), the dark matter shapes $S_{\rm dm}$ and $Q_{\rm dm}$ are in general the most predictive halo properties for gas shapes.}
    \label{fig:SHAP_mass_dependence}
\end{figure*}

In Figure \ref{fig:SHAP_mass_dependence} we show how the most predictive features change for different halo masses. For instance, we note that the most important feature in predicting gas shape in the core for $\log_{10} (M_{500c}/\Msun)\sim 12$ mass haloes is $\cvir$.  However, the relative importance of $\cvir$ decreases for increasing halo mass. Furthermore, the higher mass haloes show halo mass as the most important feature in each of the 3 considered TNG300 mass bins, consistent with the left panel from Figure~\ref{fig:shap_TNG300}. In particular, the importance of mass is strongest for haloes around $10^{13} \Msun$.  The importance of mass in the models significantly weakens for models trained using lower mass haloes from TNG100. For these lower mass haloes ($M_{500c} \leq 10^{13} \Msun$), halo formation proxies such as halo concentration and halo half-mass time are the main predictors of gas shapes in the halo core regions.

On the other hand, the right panel of Figure~\ref{fig:SHAP_mass_dependence} shows that, at all mass ranges, the outskirts gas shapes are best predicted by the underlying dark matter shape axis ratios, the $\Sdm, \Qdm$ parameters, as well as the localized dark matter densities. This result is also consistent with the results from Figure~\ref{fig:shap_TNG300}, which considers objects from a wider mass range.

These disparities in feature importance across the selected mass ranges suggest physical differences between halo cores, which are more affected by baryonic effects, and halo outskirts, which are mainly driven by accretion processes. These results confirm the power of our \pkg{XGBoost} and \pkg{SHAP} pipeline in learning connections between halo properties at a wide selection of mass and radial ranges.

Since both halo formation history and baryonic processes affect the gas distribution, the most predictive features can illuminate the relative impact of different physical processes on the halo formation and evolution. We emphasize, however, that the feature importance ranking identifies a {\it correlation} between features and gas shape at a given radius, and not {\it causation}. Despite this caveat, the correlations can be useful to help guide physical interpretations and intuition, as well as motivate further study of particular halo properties and their interdependence. For instance, $\S$\ref{sec:results_cam_comparisons} illustrates how our feature importance rankings obtained from \pkg{XGBoost} can improve the model performance of another method, CAM, by identifying a set of preferred input parameters to the CAM model.

The same machine learning pipeline can be applied to identify the most predictive  properties across simulations with different implementations of subgrid physics.  Such a comparison could inform analyses of the most interesting parameters across simulations.

\subsubsection{Limitations of SHAP}
We emphasize a limitation regarding the use of \pkg{SHAP} in this study, which may be expanded upon in future works. \pkg{SHAP} cannot be used to determine {\it causal relationships} \citep{ma2020_shap_limitations}. When applying this exercise, we need to carefully distinguish predictive power from physical effect. For instance, while we find that dark matter shapes are more predictive of gas shapes in the halo outskirts, we cannot infer which physical processes lead to this correlation from \pkg{SHAP} alone. 
The combination of \pkg{XGBoost} and \pkg{SHAP} helps in predicting values of gas shapes based on halo properties, and may lead to future scientific questions, but cannot answer these questions without a deeper analysis of the underlying physics. A more immediately useful application of our procedure would be in how the feature importance ranking identifies parameters that might be more relevant to models that can only take a limited number of input parameters for prediction.  Our analysis pipeline efficiently cuts through large data volumes to identify the most interesting features for a given task.  We further discuss this application of our approach with CAM in the next subsection.

\subsection{Comparisons with Conditional Abundance Matching (CAM)}\label{sec:results_cam_comparisons}

\begin{figure}
    \centering
    \includegraphics[width=0.47\textwidth]{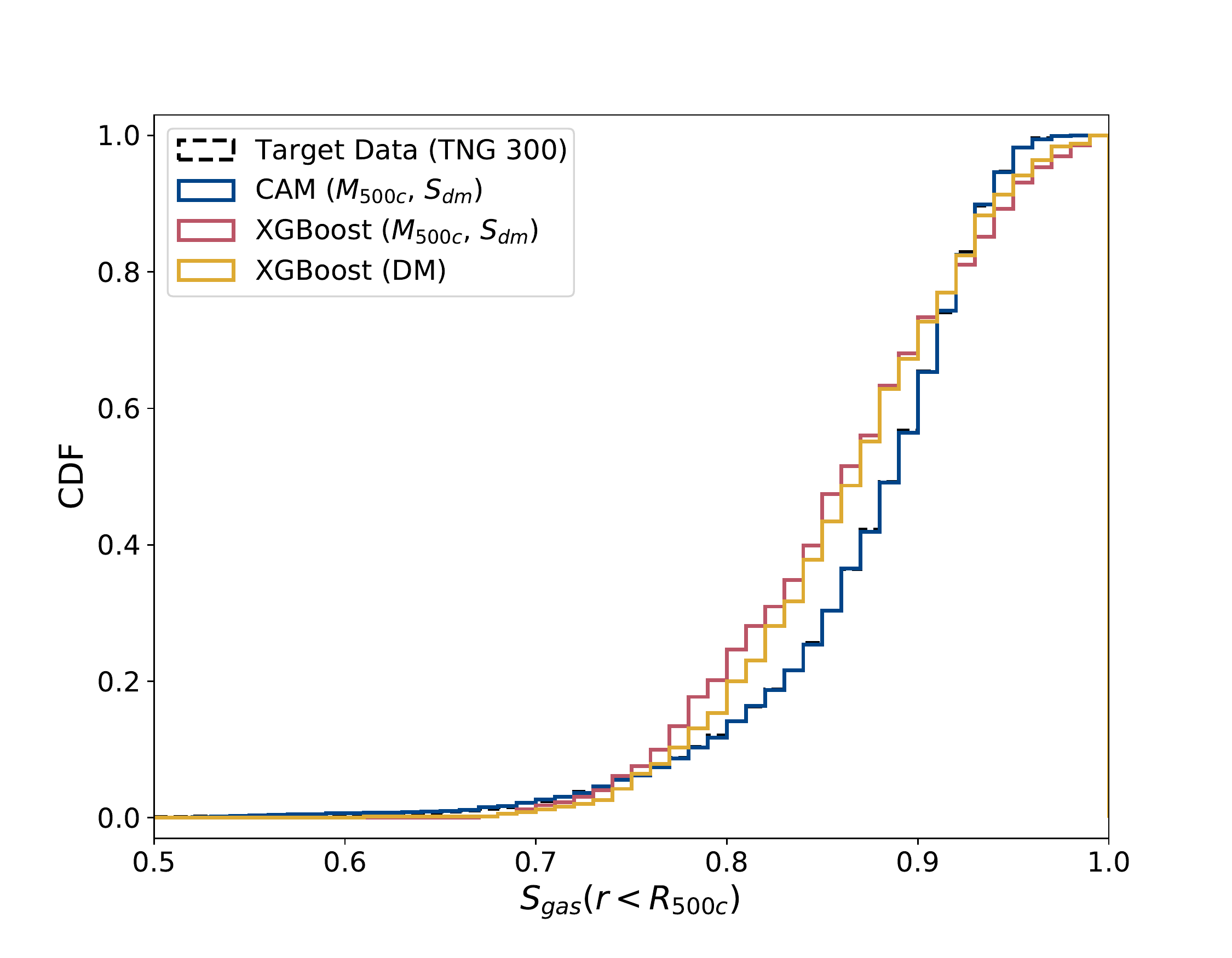}
    \caption{We compare the cumulative distributions (CDFs) of $S_{\text{gas}}(r<R_{500c})$ (which measures the shape of gas within $R_{500c}$ for each halo) between the \pkg{XGBoost} and CAM models, and the true values from TNG300. Specifically, we compare three models to the target data: (1) a CAM model obtained by using $M_{500c}$ and global $\Sdm$ as primary and secondary parameters, respectively; (2) a \pkg{XGBoost} model trained with only $M_{500c}$ and $\Sdm$ as inputs; and (3) a \pkg{XGBoost} model trained with the DM feature set, which includes all dark matter features from Table \ref{table:halo_catalog}. CAM is primarily designed to reproduce the target distribution more closely than \pkg{XGBoost}, whereas \pkg{XGBoost} is not as capable of reproducing the extrema of the target data range.}
    \label{fig:cam_xgboost_cdf}
\end{figure}

\begin{figure*}
\includegraphics[width=0.47\textwidth]{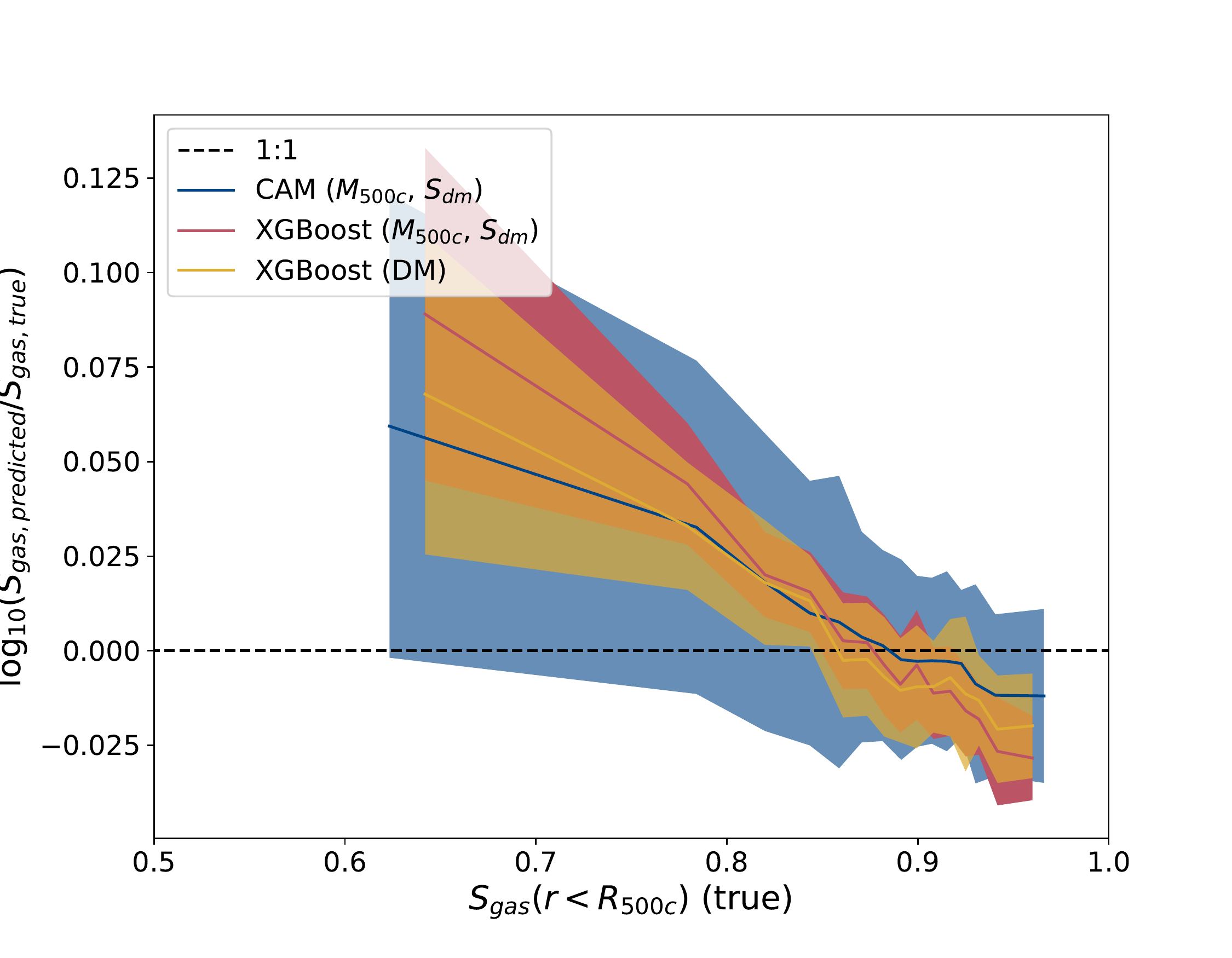}
\includegraphics[width=0.47\textwidth]{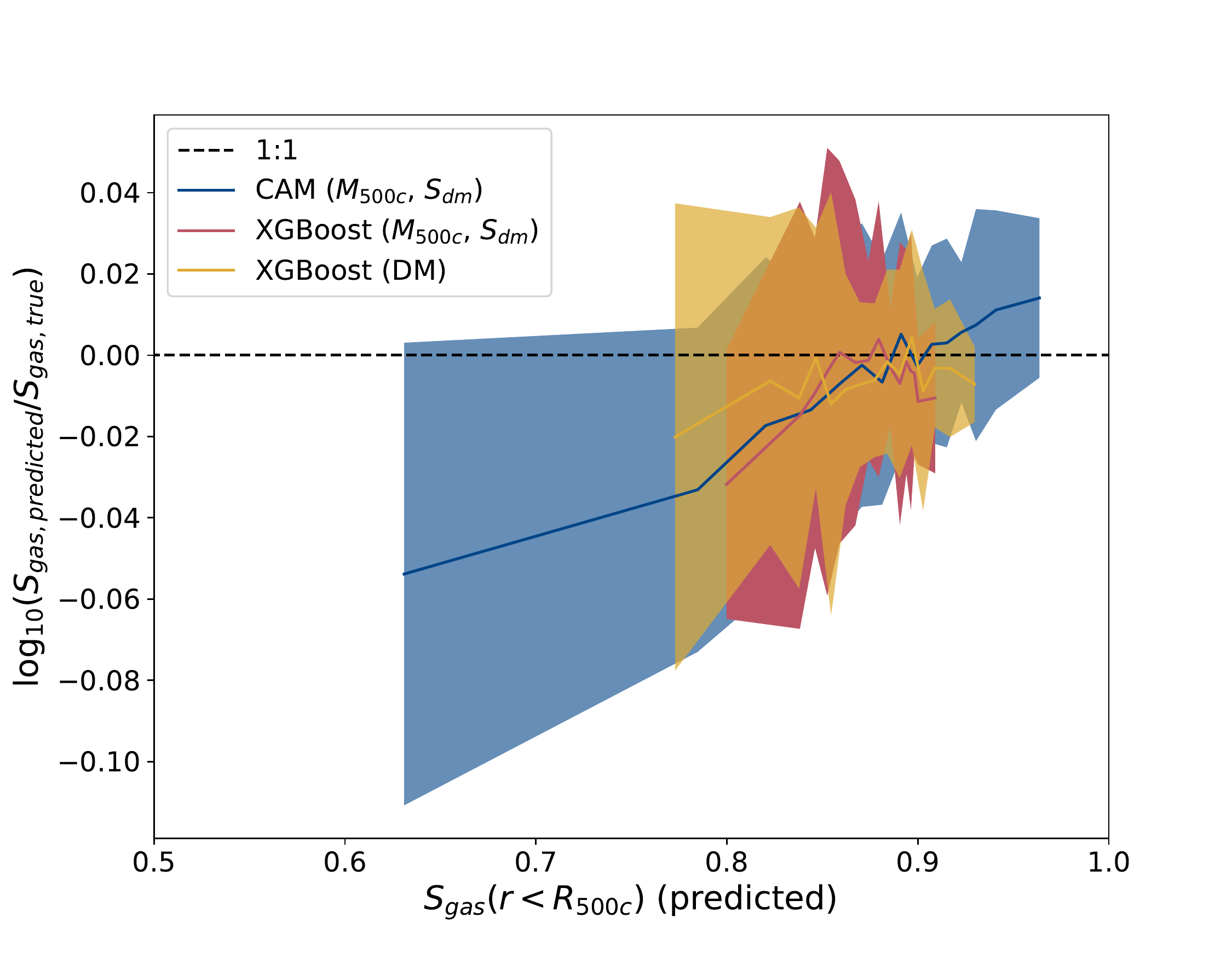}

\caption{
We compare the median and scatter of the predicted values at fixed bins of target gas shape $\Sgas$ similarly to Figure \ref{fig:predicted_vs_true_scatter_plots}, for the\pkg{XGBoost} and CAM predicted gas shapes. The {\em left} panel shows the ratio of the predicted to true $\Sgas$ values with respect to the true $\Sgas$, while the {\em right} panel shows the same ratio with respect to the predicted $\Sgas$.  The horizontal gray line indicates the ideal model. The solid lines indicate the median predicted values for each model, and the shaded regions indicate the log-normal scatter (1$\sigma$ and 2$\sigma$) around the median line. \pkg{XGBoost} outperforms CAM in both the median values and the scatter ranges, even when only trained using the same exact parameter inputs, confirming that \pkg{XGBoost} manages to learn more non-linear correlations from training data. While CAM is designed to match given distributions of data, it is not able to predict new values of a given property on a halo-by-halo basis as accurately as a machine learning method such as \pkg{XGBoost}.}
\label{fig:cam_comparisons}
\end{figure*}

\begin{table*}
\centering
\begin{tabular}{c c c c c} 
\hline\hline
Model & KS Statistic & KS p-value & Spearman Correlation & RMSE \\ [0.5ex] 
\hline
CAM ($M_{500c}$, $\cvir$) (not plotted) & $8\times 10^{-3}$ & 0.95 & 0.3 & 0.09 \\
CAM ($M_{500c}$, $\Sdm$) & $6\times 10^{-3}$ & 0.95 & 0.5 & 0.07 \\
\pkg{XGBoost} ($M_{500c}$, $\Sdm$) & 0.14 & $10^{-7}$ & 0.5 & 0.05 \\
\pkg{XGBoost} (DM) & 0.12 & $5\times 10^{-6}$ & 0.6 & 0.04 \\
\hline
\end{tabular}
\caption{Quantifying trade-offs between \pkg{XGBoost} and CAM from Figure \ref{fig:cam_comparisons}. The two-sample Kolmogorov-Smirnov (KS) statistic and p-value between the target and model CDFs from \emph{left} panel and 
Spearman correlation coefficient and RMSE of the true and predicted gas shape values from the \emph{right} panel. The KS values illustrate that CAM best reproduces the CDF, as it is designed to do.  The RMSE illustrates that \pkg{XGBoost} best captures the variance on a halo-by-halo basis.}
\label{table:cam_comparisons_table}
\end{table*}

Finally, we compare the \pkg{XGBoost} performance with an analytical method of modelling correlated halo properties. We use Conditional Abundance Matching (CAM, \citealt{Masaki_2013, Hearin_2013}) to generate another set of predicted gas shapes based on a few halo properties. Specifically, we utilize the {\tt conditional\_abunmatch} method from \pkg{halotools} \citep{hearin2017cam}. In general, CAM provides an ansatz for the dependence of a given halo observable on a primary and a secondary halo properties.  In this study, we use the halo mass $M_{500c}$ and the global dark matter shape $\Sdm$ as the respective primary and secondary halo properties. By design, CAM first matches the rank-ordered lists of $\Sgas$ and $\Sdm$ by equating the cumulative probability density distribution functions (CDF) of $\Sgas$ and of $\Sdm$ in a given $M_{500c}$ bin,
\begin{equation}
F(\Sgas | M_{500c}) = F(\Sdm | M_{500c}) ,
\end{equation}
where $F(X|Y) \equiv \int_{-\infty}^X P(x|Y)dx$ is the CDF. As CAM assumes perfect one-to-one correlation between $\Sgas$ and the secondary parameter $\Sdm$ at a given $M_{500c}$,  we reshuffle the rank-ordered list of $\Sgas$ values to match the Spearman correlation coefficient between $\Sgas$ and $\Sdm$ computed from the simulation to account for the non-perfect correlation between the target $\Sgas$ and the $\Sdm$. This procedure ensures a relation between $\Sgas$ and $\Sdm$ that closely resembles what we measure in the simulation. 
Thus, the CAM model is by nature {\em empirical}, in that it entirely depends on the input simulation (or input observation) on which it is calibrated. 

The CAM model describes the dependence of $\Sgas$ on two halo parameters. 
This is different from the \pkg{XGBoost} model, wherein we in general consider a wider set of halo parameters. To achieve a fair comparison between \pkg{XGBoost} and CAM, we use two different feature sets: the \emph{DM} feature set, combining the Dark Matter properties from Table \ref{table:halo_catalog}; and a set including only $M_{500c}$ and $\Sdm$, for a true apples-to-apples comparison. Finally, we choose to target global gas shape parameters, which we compute by including all gas cells within $R_{500c}$.  We choose this target data to compare the predictive power of CAM and \pkg{XGBoost} on a quantity that summarizes the gas distribution of the entire halo, instead of the shape measurement at specific radial bins.

We first quantify the differences in the distributions.  We perform the two-sample Kolmogorov-Smirnov (KS) test for two scenarios: (1) true TNG300 measurements compared to \pkg{XGBoost} predictions, and (2) true TNG300 measurements compared to the CAM predictions, and display the results in Table \ref{table:cam_comparisons_table}. 
The two-sample KS statistic tests whether two samples have the same distribution by comparing the maximum difference $D$ in their cumulative density distribution functions (CDF), which we show in Figure~\ref{fig:cam_xgboost_cdf}.  A smaller value of $D$ and a larger $p$-value indicate that the two samples are more likely to be drawn from the same distribution. For CAM (dark blue line), the KS statistic is $D = 8\times 10^{-3}$ (with $p = 0.95$). For \pkg{XGBoost}, $D = 0.14$ (with $p = 10^{-7}$) for the dataset with $M_{500}$ and $\Sdm$ (red line) and $D = 0.12$ (with $p = 5\times 10^{-6}$) for the full {\em DM} dataset (yellow line). The KS statistic for CAM is smaller than those for \pkg{XGBoost}. 
This illustrates that the distribution from the {\it CAM predictions better matches the distribution of the true TNG300 measurements}, the target data of the models.

Second, we compare the accuracy of gas shape predictions between CAM and \pkg{XGBoost} in Figure~\ref{fig:cam_comparisons}. The left panel shows the ratio $S_{\rm gas,true}/S_{\rm gas, predicted}$ as a function of $S_{\rm gas,true}$, with CAM predictions shown in blue and \pkg{XGBoost} in red (for the two feature model most comparable to CAM) and yellow (for the model trained with all dark matter information).  The right panel shows the same, but as a function of predicted gas values.  When we compare CAM and \pkg{XGBoost} models trained with the same two features, the scatter in \pkg{XGBoost} ($\sim 10\%$) is significantly smaller than that of CAM ($\sim 30\%$), indicating that \pkg{XGBoost} performs better than CAM in predicting $\Sgas$. At $S_{\rm gas,true}=0.4$, the \pkg{XGBoost} values are $\sim 40\%$ more accurate than those from CAM. Here,  {\it \pkg{XGBoost} outperforms CAM in model accuracy}.

Finally, CAM trained with the most important features identified from \pkg{SHAP} ($M_{500}$ and $\Sgas$) outperformed the model trained with the canonical features used by ~\citet{lau2020correlations} ($M_{500}$ and $\cvir$).  We do not include this model in Figure~\ref{fig:cam_comparisons}, since this would make the figure more difficult to read.  Instead, we summarize quantitative results in Table~\ref{table:cam_comparisons_table}.  A comparison of the first and second rows of Table~\ref{table:cam_comparisons_table} shows that the CAM model trained with features identified by \pkg{SHAP} to be important to \pkg{XGBoost} models improves across three of the metrics: KS Statistic, Spearman Correlation strength, and RMSE.  The models have equal performance quantified by the KS $\rho$-value. These results illustrate how \pkg{XGBoost} and \pkg{SHAP} can help optimize feature choices for models like CAM.

These results suggest that there are both important trade-offs and complementarity between the two methodologies. CAM can reproduce the overall target distribution, but is unable to use information from more than two features, and it is also less accurate in its predictions. \pkg{XGBoost} provides more accurate models, but cannot fully reproduce the overall target CDF and has trouble recreating the target outliers (for instance, see Figure 3 from \citealt{ntampaka_2019}). \pkg{XGBoost} also presents several advantages: it predicts target values more accurately for individual haloes; it can learn from as many halo properties as needed; it manages to account for more non-linear correlations between different features and the target data; and \pkg{SHAP} allows for interpretability of the resulting \pkg{XGBoost} models, providing feature rankings that can motivate and inform the understanding of halo formation and baryonic effects. Moreover, perhaps the most important takeaway is that \pkg{XGBoost} and \pkg{SHAP} results can motivate the use of specific features for models like CAM, optimizing the use of such methods.

\section{Conclusions}\label{sec:conclusions} 

In this work, we present interpretable machine learning models that predict gas shapes in dark matter haloes. We train predictor models with \pkg{XGBoost}, an implementation of gradient boosted trees \citep{chenandguestrin16}, with different subsets of input halo properties from the IllustrisTNG simulations. We use \texttt{TreeSHAP} implementation of \pkg{SHAP} \citep{Shapley195317AV, NEURIPS2017_7062,lundberg2020local2global}, a game-theory-based method to quantify the relative predictive power of different halo properties in the \pkg{XGBoost} models that predict gas shape at several halo radii. The main findings are:

\begin{itemize}
    \item With dark matter information only, the \pkg{XGBoost} models predict gas shapes with a mean error of $\lesssim 20 \%$ around the true $\Sgas$ values at radii $r / R_{200c} \geq 0.5$ in haloes with $M_{500c} \geq 10^{13} \Msun$. The model accuracy of dark matter only predictions of gas shape improves towards larger radii.  The predictive power reflects that accretion processes driven by the underlying dark matter distribution more heavily affect gas shapes in the outer regions of haloes, whereas baryonic effects fundamentally impact the gas shapes in the halo cores (See Figures \ref{fig:predicted_vs_true_scatter_plots} and \ref{fig:rmses_feature_sets_ratios_test}).
    
    \item For TNG300 groups and clusters ($M_{500c} \geq 10^{13} \Msun$), \pkg{SHAP} identifies halo mass as the most important predictor of gas shapes at the core ($r/R_{200c} \leq 0.2$), and dark matter shape is the main predictor in the outer regions ($r/R_{200c} \geq 0.2$) of the haloes (See Figure \ref{fig:shap_TNG300}).
    
    \item While the \pkg{XGBoost} and \pkg{SHAP} pipeline can account for trends across a wide range of halo masses, this pipeline is also powerful when identifying how the importance of different features vary with halo mass. We trained models to predict gas shapes for haloes within narrower mass bins and identified systematic changes in \pkg{SHAP} feature rankings.  For lower mass haloes (up to $10^{13} \Msun$), halo formation proxies (e.g. halo concentration $\cvir$) are the main predictor of gas shapes at the halo cores. In more massive haloes, halo mass is the strongest predictor (see Figure~\ref{fig:SHAP_mass_dependence}). On the other hand, dark matter shapes and densities are the most important features for models predicting gas shapes in the outskirts, regardless of halo mass.
    
    \item We compare \pkg{XGBoost} with an alternative model, CAM (see Figure \ref{fig:cam_comparisons}).  While CAM, by construction, reproduces the cumulative distribution function of gas shapes {(see Figure~\ref{fig:cam_xgboost_cdf}), \pkg{XGBoost} generates more accurate predictions of gas shapes in the sampled haloes with the same input features to both models, as \pkg{XGBoost} captures gas shape dependence of more than one halo properties compared to CAM.  
    
    \item We emphasize that results from our \pkg{XGBoost} and \pkg{SHAP} pipeline can improve the application of other methods, such as} CAM. CAM model predictions improve across all metrics when using features identified from the \pkg{XGBoost} and \pkg{SHAP} pipeline ($M_{500c}$ and $\Sdm$), compared with the canonical features used for CAM ($M_{500c}$ and $\cvir$). (See Table~\ref{table:cam_comparisons_table}).
    \end{itemize}

While we have demonstrated that \pkg{XGBoost} and \pkg{SHAP} are able to reproduce gas shapes in dark matter haloes, there are limitations to this approach. First, common to all machine learning methods, the predicted power of the model is limited by the input training sample. Extrapolating gas shape predictions in haloes that are not represented in the input simulation sample are not accounted for will lead to biases. In addition, while \pkg{XGBoost} and \pkg{SHAP} highlight the correlations between gas shape and other physical properties of the halo, they do not provide the underlying causal connections. A full understanding of the physics that governs gas shapes in haloes will require complementary approaches with e.g., analytic modeling. 

To our knowledge, the current work demonstrates one of the first applications of {\em interpretable} machine learning techniques (with \pkg{XGBoost} and \pkg{SHAP}) in connecting baryons with the underlying dark matter haloes.  We specifically focus on understanding the physical modelling of the distribution of the hot X-ray emitting gas in galaxy groups and clusters.  By applying machine learning techniques to the outputs of modern hydrodynamical cosmological simulations, we identified important physical properties that predict gas shapes at different radii in a dark matter halo. In general, this approach can be extended to other halo properties, allowing us to improve empirical models like CAM. 

This work demonstrates that insights provided by interpretable machine learning approach can advance a physically-motivated and computationally efficient halo models for upcoming multi-wavelength cosmological surveys. 

\section*{Acknowledgements}
We thank Dhayaa Anbajagane, Han Aung, Andrew Hearin, Phil Mansfield, and Michelle Ntampaka for their constructive comments that improved the contents and presentation of the paper. 
CA was supported by the Leinweber Center for Theoretical Physics and the LSA Collegiate Fellowship at the University of Michigan. DB acknowledges support through NSF grant AST-1814053. AF is supported by a Michigan Institute for Data Science (MIDAS) Fellowship. EL acknowledges the University of Miami for support. We gratefully acknowledge use of the Harvard Odyssey and Yale Grace clusters as part of this work. 

\section*{Data Availability}
The TNG simulation data used in this paper is publicly available on \url{https://www.tng-project.org}. The scripts used to analyze the data and generate the plots are available upon request.

\appendix

\section{XGBoost Performance for all Feature Subsets}\label{sec:appendix_rmse}

This Appendix displays the entire result set obtained from all 175 \pkg{XGBoost} models described in $\S$ \ref{sec:featuresubsets}. We expand on Figure \ref{fig:rmses_feature_sets_ratios_test} and include additional lines, which correspond to different feature subsets applied when training the \pkg{XGBoost} models.

\begin{figure*}
\includegraphics[width=\textwidth]{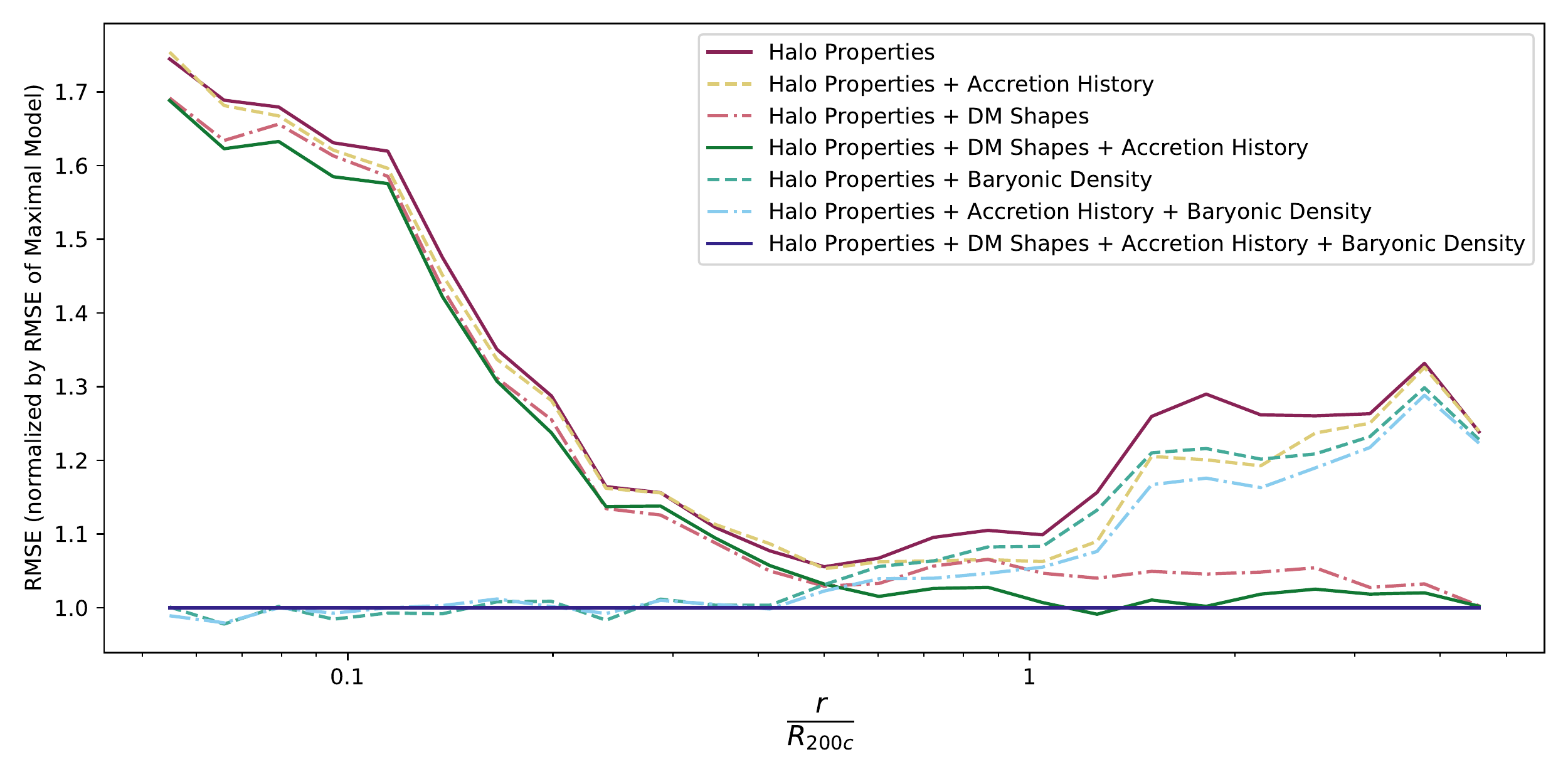}

\caption{Performance of \pkg{XGBoost} models including different feature subsets, trained for gas shapes measured at different radial bins. We normalize the RMSE values by those of the model including all available features, i.e., the {\em maximal model}. The performance towards inner radial bins confirms the importance of Baryonic Properties when predicting gas shapes at the halo cores. Additionally, there is a substantial improvement caused by including Dark Matter Shapes into the models at larger radii, showing the importance of including Dark Matter Shapes when predicting gas shapes at the halo outskirts.}

\label{fig:appendix_rmses}
\end{figure*}

The main goal of this work involves showcasing the relative impact of different features in the final performance of the \pkg{XGBoost} models. To facilitate such relative comparisons, we display the relative RMSE values of each \pkg{XGBoost} model, instead of the absolute RMSE values. Figure \ref{fig:appendix_rmses} shows the performance of each \pkg{XGBoost} model, {\em normalized} by the performance of the maximal model, i.e., the model including all available features.

Figure \ref{fig:appendix_rmses} supports the results mentioned in $\S$ \ref{sec:featuresubsets}. In particular, it confirms that, in the inner radial bins, the model RMSE improves substantially (~70\%) with the inclusion of Baryonic Properties. Furthermore, it shows the impact of including Dark Matter Shape profiles in the outermost radial bins, improving the models' RMSEs by ~30\%.

\section{Comparison between TNG100 and TNG300 haloes at the same mass bin}\label{sec:appendix_100_300_comparison}

We perform a \pkg{XGBoost} and \pkg{SHAP} analysis of haloes at several mass bins, to determine the mass dependence of the feature importance rankings. Since \pkg{XGBoost} generates more accurate models when trained with larger datasets, we require haloes from both TNG100 and TNG300 to obtain large enough datasets for all desired mass ranges. However, this combination of haloes from two simulations is \emph{a priori} not valid, since in principle \pkg{XGBoost} and \pkg{SHAP} could behave differently depending on the simulation originating the halo samples. As a result, to validate the comparison performed in \ref{sec:results_shap_mass_dependence}, we first verify that using both TNG100 and TNG300 datasets is a fair combination.

\begin{figure*}
\includegraphics[width=0.47\textwidth]{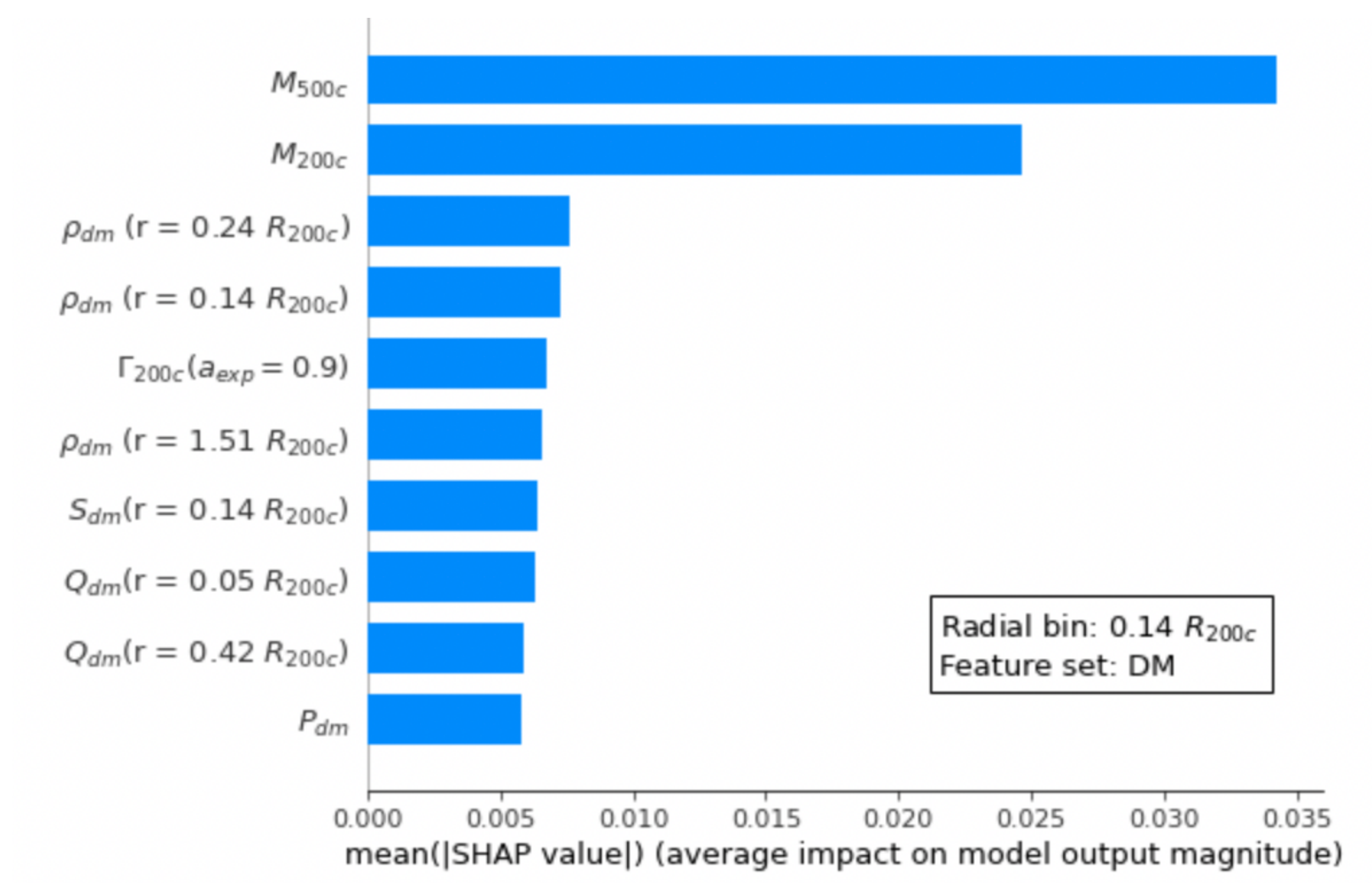}
\includegraphics[width=0.47\textwidth]{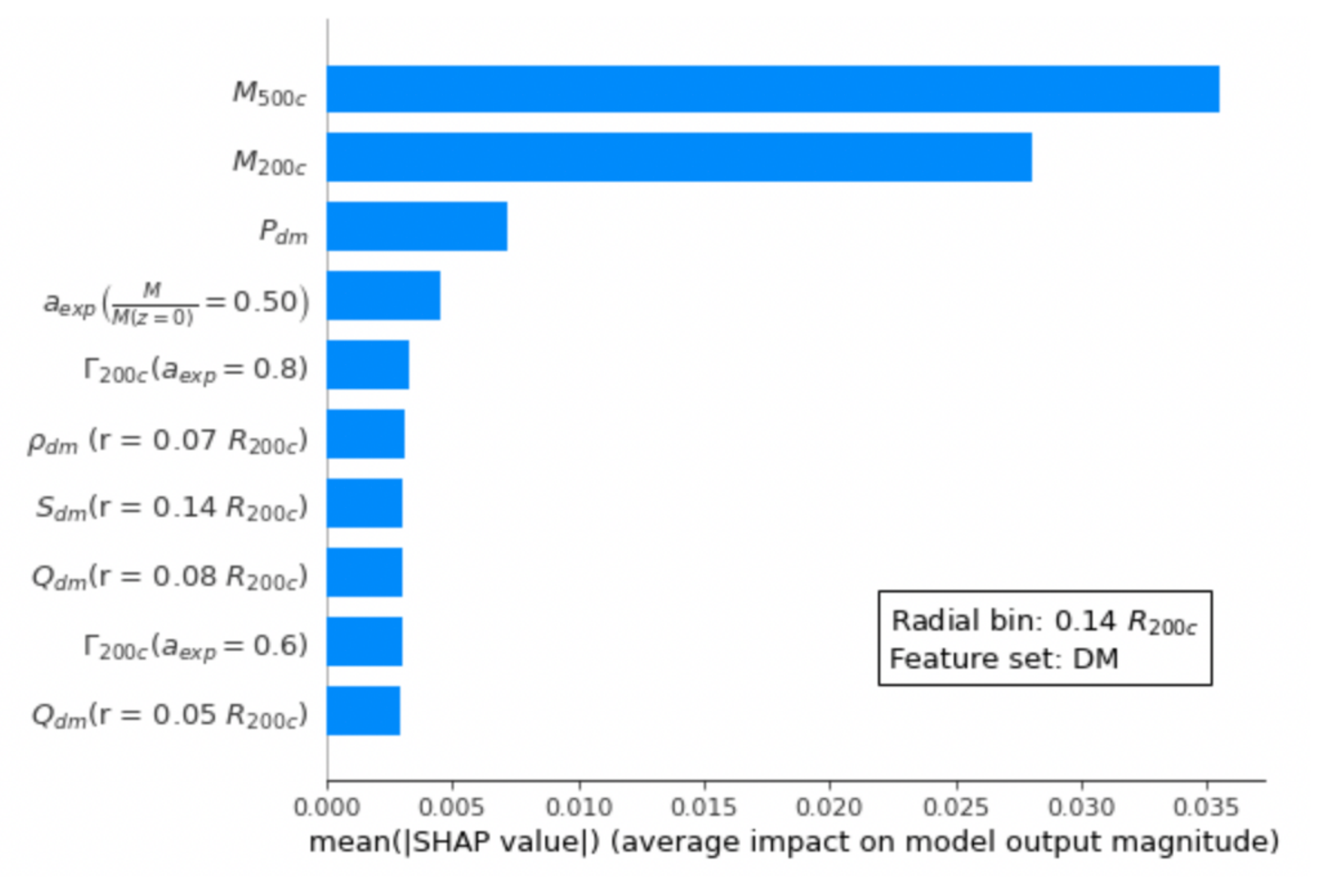}
\includegraphics[width=0.47\textwidth]{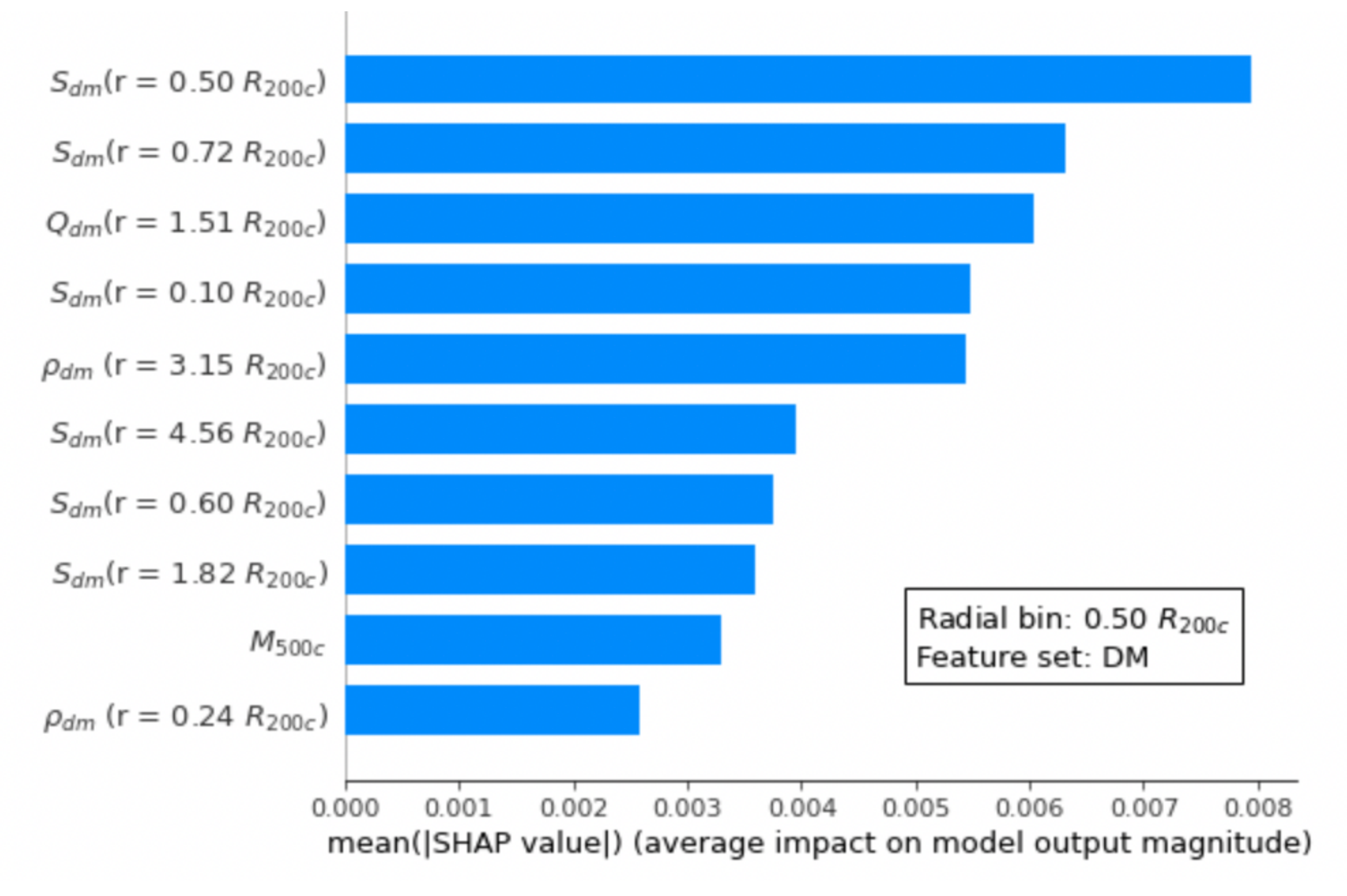}
\includegraphics[width=0.47\textwidth]{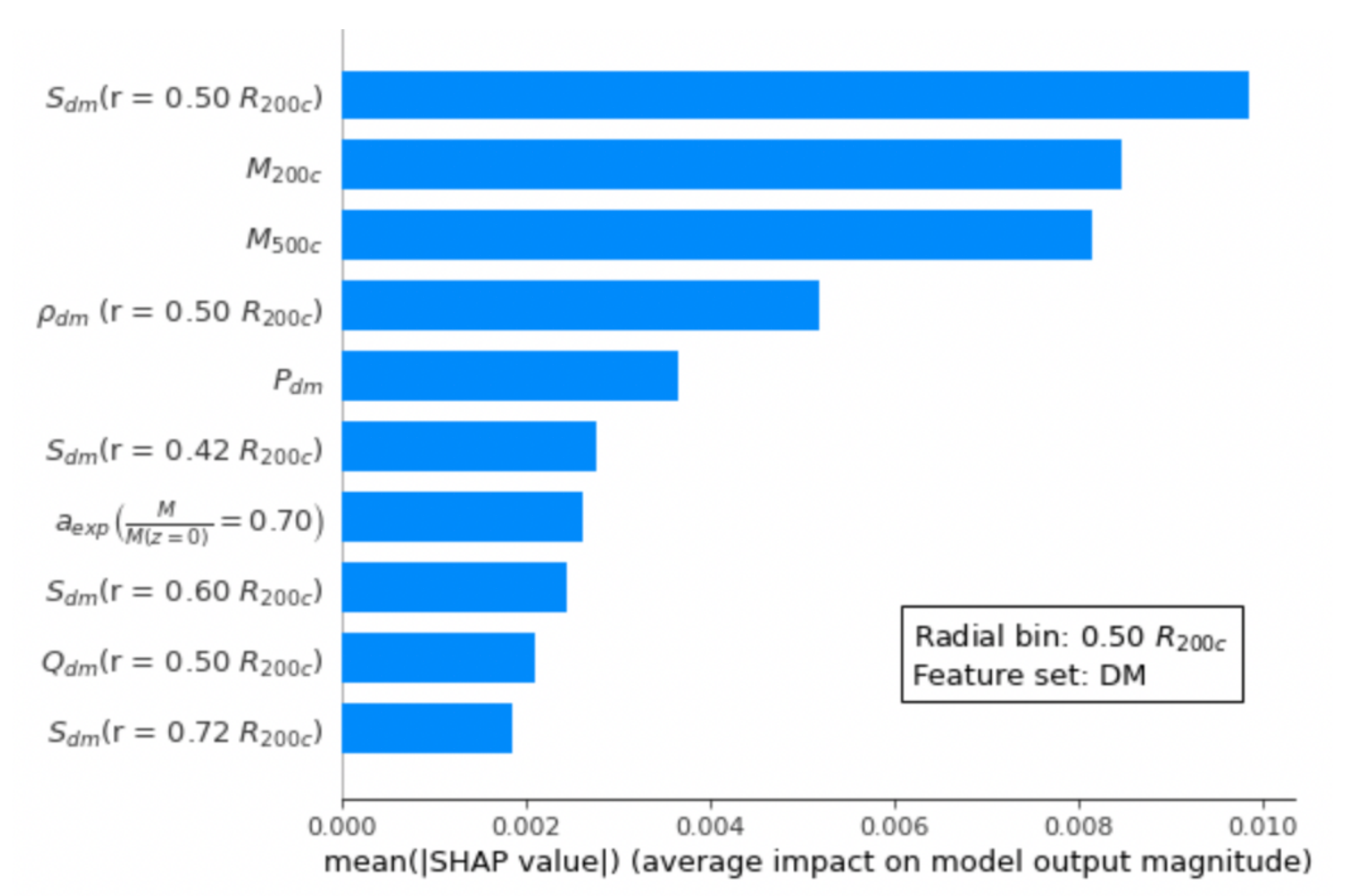}
\includegraphics[width=0.47\textwidth]{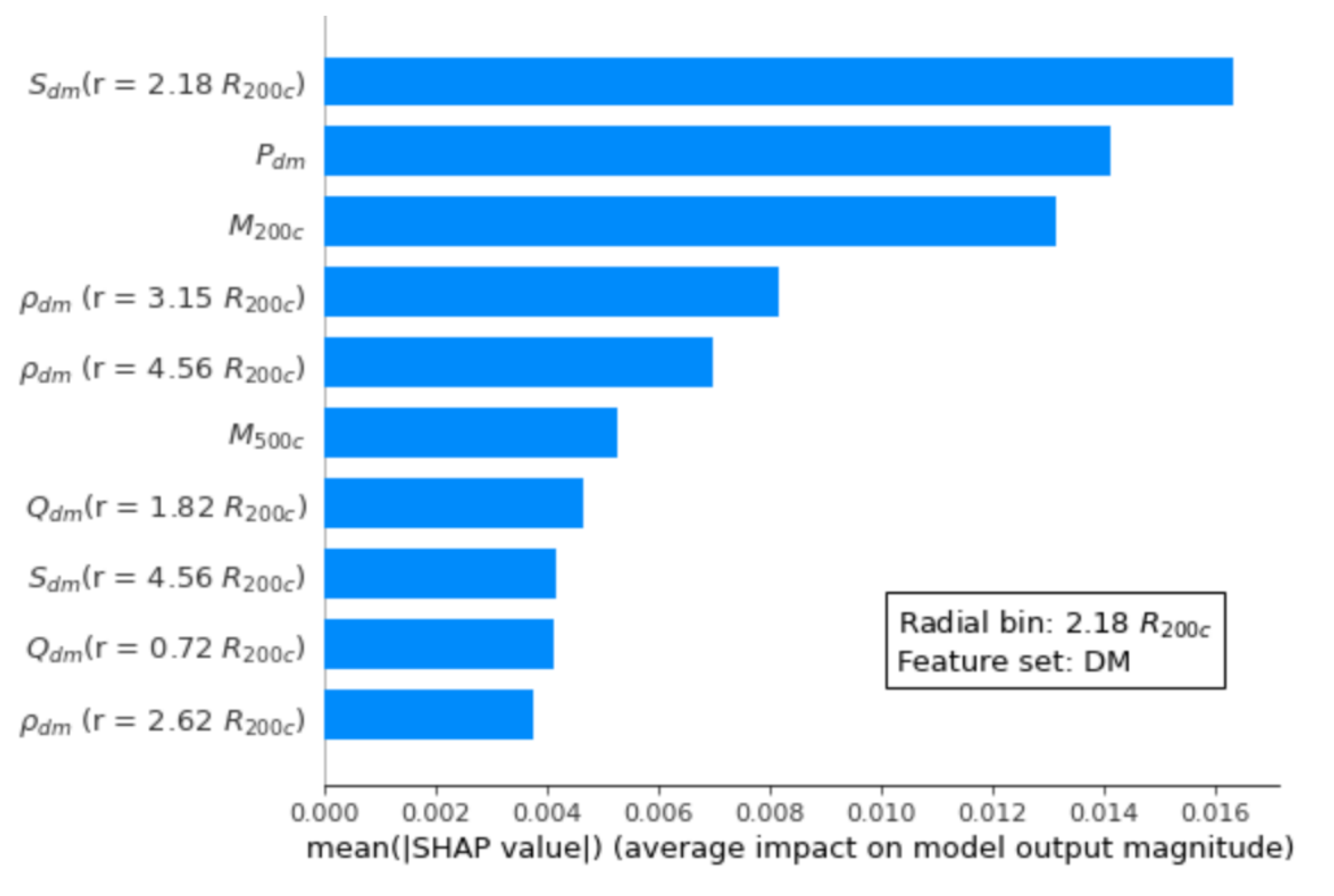}
\includegraphics[width=0.47\textwidth]{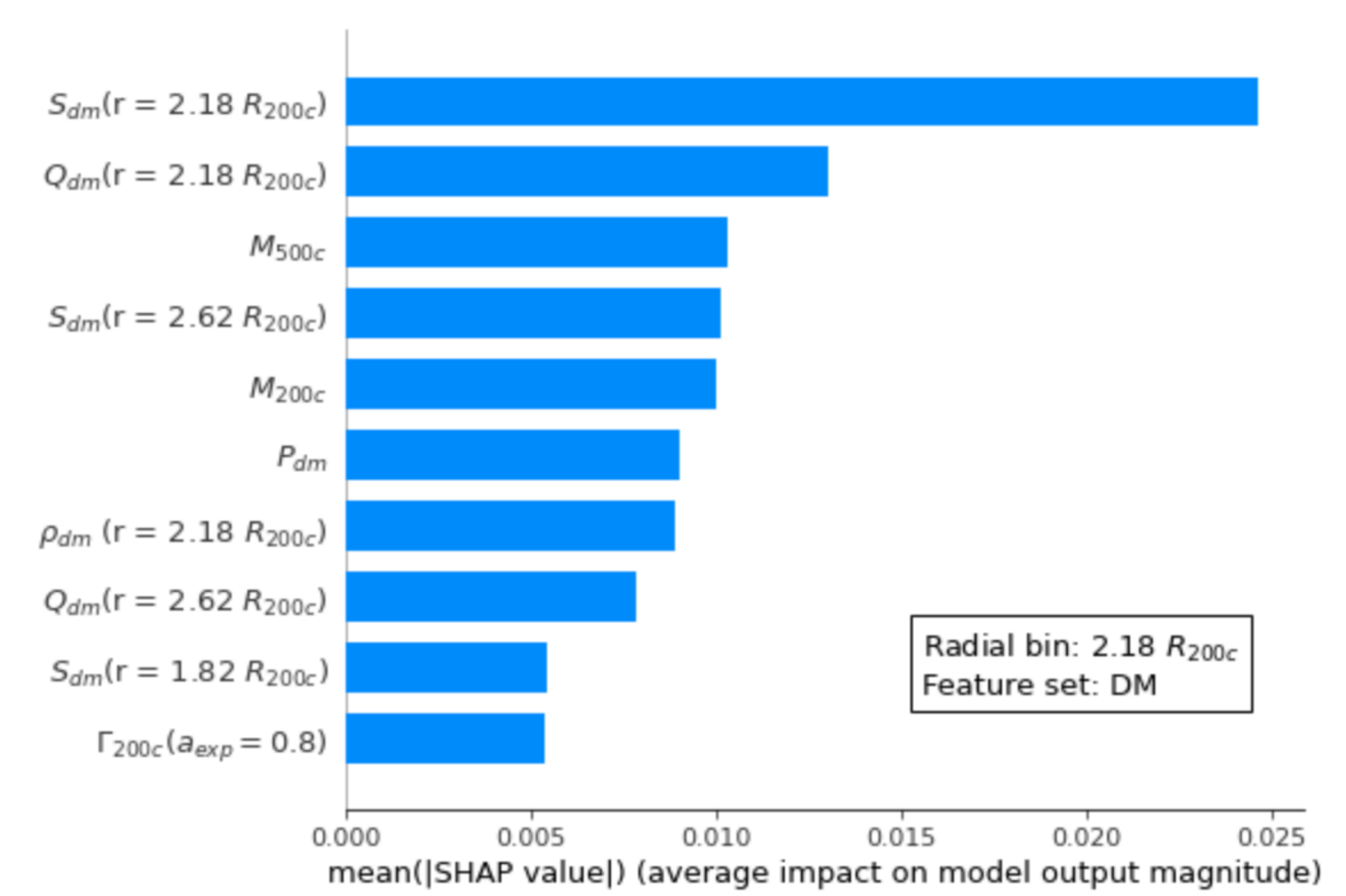}

\caption{Comparison of \pkg{SHAP} feature importance rankings for TNG100 (\emph{left panel}) and TNG300 (\emph{right panel}) haloes. We select haloes within $10^{13} \leq M_{500c}/\Msun < 10^{14}$, to establish a fair comparison between \pkg{SHAP} results across simulations. We compare models trained to predict gas shapes at 3 radial bins, namely inner (\emph{top panels}), intermediate (\emph{middle panels}), and outer (\emph{bottom panels}) bins (see \ref{sec:results_distributions} for more detailed descriptions of these bins). 
We note that, at every radial bin, the feature importance rankings are mostly similar, with the disparities caused by relatively small differences in the mean \pkg{SHAP} values. This equality indicates that \pkg{SHAP} is robust across the TNG100 and TNG300 simulations, and can hence be applied to compare haloes in both simulations, which is done in \ref{sec:results_shap_mass_dependence}.}
\label{fig:appendix_tng100_tng300_comparison}
\end{figure*}

To enable a fair comparison of haloes across mass ranges, we first create a fair comparison of \pkg{XGBoost} and \pkg{SHAP} between TNG100 and TNG300. We control for the mass range, and select haloes with $10^{13} \leq M_{500c} < 10^{14}$ for this comparison. For \pkg{SHAP} to be considered robust across simulations, it should in principle result in similar feature rankings for both TNG100 and TNG300 halo samples. In fact, Figure \ref{fig:appendix_tng100_tng300_comparison} shows that, within the error for \pkg{SHAP} values, the rankings for both halo samples are equal.

This result confirms that the \pkg{XGBoost} and \pkg{SHAP} pipeline developed in this study can handle different simulations, and (provided the simulations themselves use similar physical models) will result in similar feature importance rankings in a simulation-agnostic manner. Furthermore, this result enables the wide mass range comparison performed in \ref{sec:results_shap_mass_dependence}, and allows the pipeline to be applied for both galaxy-sized and cluster-sized groups.

\bibliographystyle{mnras}
\bibliography{references}

\bsp	
\label{lastpage}
\end{document}